\documentclass[acmtog]{acmart}
\acmSubmissionID{2494}

\usepackage{booktabs} %

\usepackage{bibunits}

\usepackage{stfloats}
\usepackage{subcaption}

\usepackage[ruled]{algorithm2e}

\SetAlFnt{\small}
\SetAlCapFnt{\small}
\SetAlCapNameFnt{\small}
\SetAlCapHSkip{0pt}

\setcopyright{acmlicensed}
\acmJournal{TOG}
\acmYear{2025} \acmVolume{44} \acmNumber{6} \acmArticle{233} \acmMonth{12}\acmDOI{10.1145/3763367}

\usepackage{enumitem}

\begin{document}
\title{Learning to Ball: Composing Policies for Long-Horizon Basketball Moves}

\author{Pei Xu}
\orcid{0000-0001-7851-3971}
\affiliation{%
 \institution{Stanford University}
 \country{USA}}
\email{peixu@stanford.edu}

\author{Zhen Wu}
\orcid{0009-0006-7800-9314}
\affiliation{%
 \institution{Stanford University}
 \country{USA}}

\author{Ruocheng Wang}
\orcid{0009-0005-8404-6405}
\affiliation{%
 \institution{Stanford University}
 \country{USA}}

\author{Vishnu Sarukkai}
\orcid{0009-0006-9809-9994}
\affiliation{%
 \institution{Stanford University}
 \country{USA}}

\author{Kayvon Fatahalian}
\orcid{0000-0001-8754-0429}
\affiliation{%
 \institution{Stanford University}
 \country{USA}}

\author{Ioannis Karamouzas}
\orcid{0009-0000-4315-6556}
\affiliation{%
 \institution{University of California, Riverside}
 \country{USA}} 

\author{Victor Zordan}
\orcid{0000-0002-7309-7013}
\affiliation{
 \institution{Roblox}
 \country{USA}}
\affiliation{%
 \institution{Clemson University}
 \country{USA}}

\author{C. Karen Liu}
\orcid{0000-0001-5926-0905}
\affiliation{%
 \institution{Stanford University}
 \country{USA}
}

\begin{abstract}
Learning a control policy for a multi-phase, long-horizon task, such as basketball maneuvers, remains challenging for reinforcement learning approaches due to the need for seamless policy composition and transitions between skills. A long-horizon task typically consists of distinct subtasks with well-defined goals, separated by transitional subtasks with unclear goals but critical to the success of the entire task. Existing methods like the mixture of experts and skill chaining struggle with tasks where individual policies do not share significant commonly explored states or lack well-defined initial and terminal states between different phases.
In this paper, we introduce a novel policy integration framework to enable the composition of drastically different
motor skills in multi-phase long-horizon tasks with ill-defined intermediate states. %
Based on that, we further introduce a high-level soft router to enable seamless and robust transitions between the subtasks.
We evaluate our framework on a set of fundamental basketball skills and challenging transitions.
Policies trained by our approach can effectively control the %
simulated character to interact with the ball and accomplish the long-horizon task specified by real-time user commands, without relying on ball trajectory references. %
\end{abstract}



\begin{teaserfigure}
\centering
    \includegraphics[width=\linewidth]{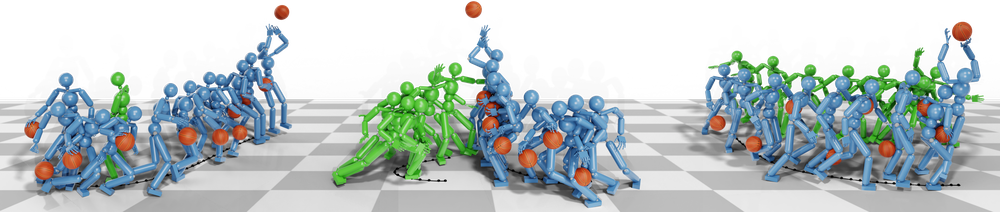}
    \caption{We introduce a novel policy integration framework to enable the composition of drastically different motor skills in multi-phase, long-horizon tasks, among them, shoot-off-the-dribble, catch-and-shoot, and board-and-bang (grabbing an offensive rebound and scoring immediately).
    }
 \label{fig:teaser}
\end{teaserfigure}

\maketitle

\section{Introduction}

Many real-world tasks consist of complex objectives that can be broken down into sequences of differing subtasks. Successfully executing these multi-phase, long-horizon tasks demands the mastery of heterogeneous skills and the ability to transition seamlessly between them. Basketball provides a compelling example of these challenges. For example, a fundamental maneuver, ``shoot-off-the-dribble'', involves distinct subtasks such as dribbling, gathering the ball and shooting, as well as the ability to transition between these skills, ultimately culminating in the ball successfully going into the hoop.  However, while \emph{dribble} and \emph{shoot} are characterized by well-defined stand-alone goals, \emph{gather} acts largely as a transition subtask with poorly defined starting and ending states.  Thus, executing such multi-phase tasks challenges control methods proposed to date.

Reinforcement learning (RL) has shown promise in training policies for individual skills  for physics-based character control~\cite{peng2018deepmimic,shi2023hci,chentanez2018physics,yin2021discovering,kwiatkowski2022survey},
but composing these policies into a cohesive framework remains an open problem. Previous approaches have attempted to address this through methods like a mixture of experts~\cite{scadiver2,peng2018sfv}. However, this technique relies on sufficiently exploring shared states across individual policies—a condition that does not hold for tasks like dribbling and shooting. Another line of work known as skill chaining~\cite{konidaris2009skill,lee2021adversarial,chen2023sequential,liu2017learning,clegg2018learning} allows concatenation of policies but requires each skill to have a well-defined set of terminal states. This limitation renders it ineffective for intermediate tasks where the subtask's goal depends on the context of the subsequent policy. For example, the gathering motion between dribbling and shooting can be intuitively described as “bringing the agent to a state where shooting is possible.” However, crafting a reward function based solely on state or action variables to reflect this goal is challenging.

To tackle the problem of building policies for ill-defined, intermediate subtasks, we introduce a policy integration method to compose drastically different and/or ill-defined skills to achieve a multi-phase, long-horizon task. The core idea is to first train policies for well-defined subtasks independently and then use these policies to guide the training of the ambiguous, intermediate subtasks. Specifically, for a task sequence consisting of subtasks A, B, and C, where A and C have well-defined task goals but B does not, we train B using policy A to define a valid initial state distribution and policy C’s state value function to shape the terminal reward. To further improve the transition between B and C, we simultaneously adapt the pretrained policy C to the states generated by B under training. In this process, a state value estimator optimized in tandem with the adapted C will be provided to reflect the up-to-date state value evaluation for policy B optimization.
With the primitive policies for all subtasks in place, we finally train a high-level soft-routing policy that directs the execution of those primitive policies based on real-time external commands, such as dribbling destination
and velocity, or a jump-shot action. %

Another challenging aspect in learning policies for multi-phase, long-horizon tasks is the heterogeneity of movement that demands diverse and extensive human motion data. Previous work on basketball motion synthesis has demonstrated compelling results when using structured data with corresponding full body, fingers, and basketball movements for physics-based character control~\cite{liu2018learning,wang2024skillmimic,wang2023physhoi,starke2020local}. However, such a special dataset is hard to scale for training a general policy capable of performing under a wide array of conditions.
In our work, instead, we demonstrate the generation of policies from unstructured data.  We leverage a diverse collection of basketball motion data, including full-body motions without hands, and hand-only motions, as well as motion examples from unstructured videos. 
To enrich locomotion behaviors, some normal running motions are also included.
Our method makes no assumptions about the correspondence across datasets or availability of ball trajectories.

Our results show that the proposed method enables the agent to perform smooth and coordinated basketball maneuvers, from gross body movements to fine finger actions, while responding adaptively to user commands. The agent can freely play basketball in real-time—for instance, dribbling to any location at variable speeds and finishing with a jump shot from any direction, achieving a shooting accuracy of 91.8\% on a professional court. We further demonstrate team play with multiple agents interacting through catching, passing, rebounding, and defending. Extensive ablation studies validate key design choices, such as soft routing and policy fine-tuning, and expose the limitations of existing methods in handling ambiguous subtasks. By addressing skill integration and phase transitions in long-horizon tasks, our approach advances the capabilities of reinforcement learning in dynamic, interactive environments.

\section{Related Work}

Creating robust controllers for long-horizon, multi-phase tasks remains a core challenge in reinforcement learning (RL) and character animation. 
Our work mainly builds on two areas of research: deep RL for physics-based character control and policy composition for executing multi-phase tasks.

Traditional approaches of physics-based character control, including dexterous control, typically rely on trajectory optimization and/or manually designed heuristic rules to achieve control in a planning or classic optimization way~\cite{liu2008synthesis,liu2009dextrous,mordatch2012contact,wang2013video,ye2012synthesis,chen2023synthesizing}.
Early work also explored using pre-collected mocap~\cite{pollard2005physically,kry2006interaction,zhao2013robust} to generate human-like motions through imitation.
During recent years, imitation learning using deep RL for policy optimization has drawn wide attention and become a popular approach for physics-based character control~\cite{peng2018deepmimic,peng2021amp,peng2022ase,motionvae,controlvae,gail2,iccgan,yao2023moconvq,scadiver2,zhu2023neural,composite}.
The deep RL framework has demonstrated remarkable success in training policies for physics-based character control across diverse domains, including locomotion~\cite{peng2017deeploco,xie2020allsteps}, racket sports~\cite{zhang2023learning,wang2024strategy}, ,
ball games~\cite{liu2022motor,liu2018learning,wang2024skillmimic,kim2025physicsfc}, 
instrument performance~\cite{guitar,piano,robopianist2023,luo2024learning}
and object manipulation~\cite{yang2022learning,xie2023hierarchical,bae2023pmp,wu2024human}. 
Our approach follows the recent paradigm of adversarial imitation learning~\cite{iccgan,peng2021amp} %
combining a GAN-like architecture with reinforcement learning for motion imitation with goal-directed control,
and perform motion synthesis given partially observable reference motions collected from multiple disparate sources.
Policies trained with our approach enable the character to interact with the basketball validly in a human-like manner,
without ball trajectories references.

Long-horizon strategic behaviors in real life, like basketball playing, often involve multiple-phased subtasks, and demand the executor to effectively chain a bunch of distinct primitive skills into a coherent sequence for task execution.
To chain multiple primitive policies,
traditional methods often assume that two consecutive skills will share some common states in which the succeeding policy can take over the character~\cite{iccgan,pan2024synthesizing,wang2024sims,wang2024skillmimic,liu2017learning,xiao2023unified}.
Recent research focuses on aligning skill transitions through state distribution matching. Techniques include regularizing terminal distributions~\cite{lee2021adversarial}, modifying initial state distributions~\cite{konidaris2009skill}, or employing bi-directional optimization to iteratively refine both~\cite{chen2023sequential}.
While these methods improve robustness, they struggle when state distributions between skills are ill-defined.
For example, in the shoot-off-the-dribble task, dribbling and shooting are subtasks with clearly defined goals, but may have completely disjoint in-betweening states.
The intermediate ball-gather behaviors are needed to close the gap.
However, there is no clear phase division for dribbling and shooting, and, thereby, we cannot train a ball-gather policy simply in a standalone way.
In this work, we present an approach to achieve such intermediate policies with ill-defined initial and terminal states.
Based on that, a high-level
policy is introduced to perform policy composition hierarchically for seamless transitions between primitive policies.

Previous literature explored hierarchical reinforcement learning for planning tasks~\cite{kulkarni2016hierarchical,vezhnevets2017feudal,peng2017deeploco}, where high-level controllers generate goals for low-level controllers to execute,
while in this work, we focus on composing multiple policies 
in a mixture-of-experts style
through a hierarchical architecture.
Early work uses hierarchical architectures to compose multiple primitive policies/poses through weighted averaging in single-phase tasks~\cite{MCPPeng19,ranganath2019low},
or through hard routing~\cite{tessler2017deep,wang2024skillmimic,wang2024strategy} to pick a primitive policy from a pre-trained policy set for control at each moment.
To generate human-like motions, the former approach typically needs additional imitation learning during composition to avoid unnatural behaviors caused by too much averaging.
The latter one, on the other hand, would suffer the challenge of policy transition when switching between different primitive policies.
Our proposed soft-routing scheme allows policy composition by weighted averaging, and, meanwhile, encourages one primitive policy to dominate the control at each moment, thereby ensuring seamless and natural transition between heterogeneous policies.

\section{Method Overview}\label{sec:overview}

\begin{figure}
    \centering
    \includegraphics[width=.95\linewidth]{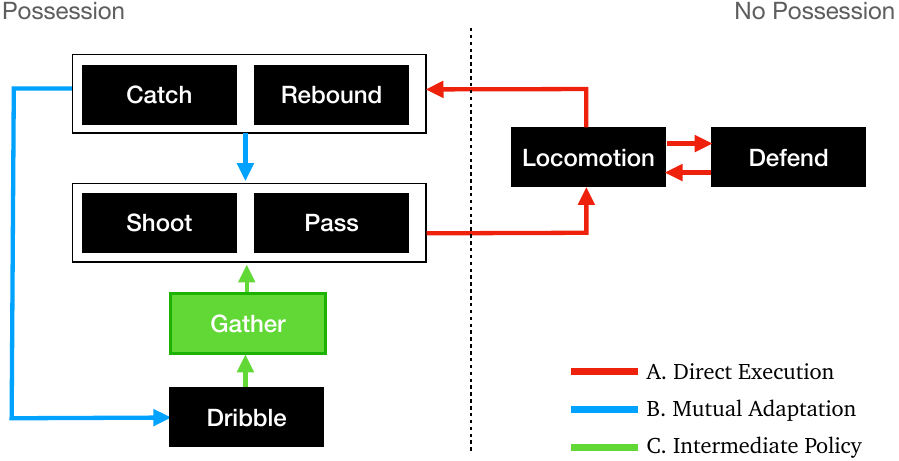}
    \caption{Our physically simulated character is capable of performing seven distinctive basketball skills and transitioning between them. The policies (shown as black boxes) include both offensive and defensive skills. The transitions between subtasks can be categorized into three types ordered by increasing difficulty: A. \textbf{Direct Execution}; B. \textbf{Mutual Adaptation}; and C. \textbf{Intermediate Policy}. The most challenging case, Type C, requires an additional intermediate policy (shown as a green box) to be trained to bridge the gap between two otherwise incompatible policies.}
    \label{fig:transition_diagram}
\end{figure}

We present a novel method for composing RL policies across distinct subtasks, enabling a physically simulated character to perform long-horizon, complex tasks such as playing basketball. A proficient basketball player must execute a wide range of fundamental skills and, importantly, transition fluidly between them without losing control of the ball. While the complete repertoire of basketball skills is extensive, we select a core set of seven skills and their transitions to showcase our method's ability to generate physically simulated players capable of unscripted, continuous, and interactive basketball behaviors~(Figure~\ref{fig:transition_diagram}).
The main focus of this work is to enable seamless transitions between subtasks. We categorize these transitions into three types, ordered by increasing difficulty. This hierarchy also guides our approach when selecting the appropriate transition method for a given pair of policies:
\begin{enumerate}[label=\Alph*.]
    \item \textbf{Direct Execution}: Succeeding policy can be executed directly from terminal state of the preceding policy. Used when transitioning between behaviors when 
    the two consecutive policies share common states generally, such as %
    \emph{Shoot} to \emph{Locomotion} or \emph{Locomotion} to \emph{Defend}.
    \item \textbf{Mutual Adaptation}: Succeeding policy must adapt to novel initial states produced by the preceding policy while the preceding policy must lead to a state manageable by the succeeding policy. For instance, transitioning from \emph{Catch} to \emph{Shoot} may fail under direct execution if the ball is not in a familiar configuration for the shooting policy.
    \item \textbf{Intermediate Policy}: This transition requires an intermediate policy to bridge incompatible subtasks. For example, transitioning from \emph{Dribble} to \emph{Shoot} or \emph{Pass} often demands a \emph{Gather} policy to reposition the ball appropriately.
\end{enumerate}

\begin{figure*}[t]
    \centering
    \includegraphics[width=\linewidth]{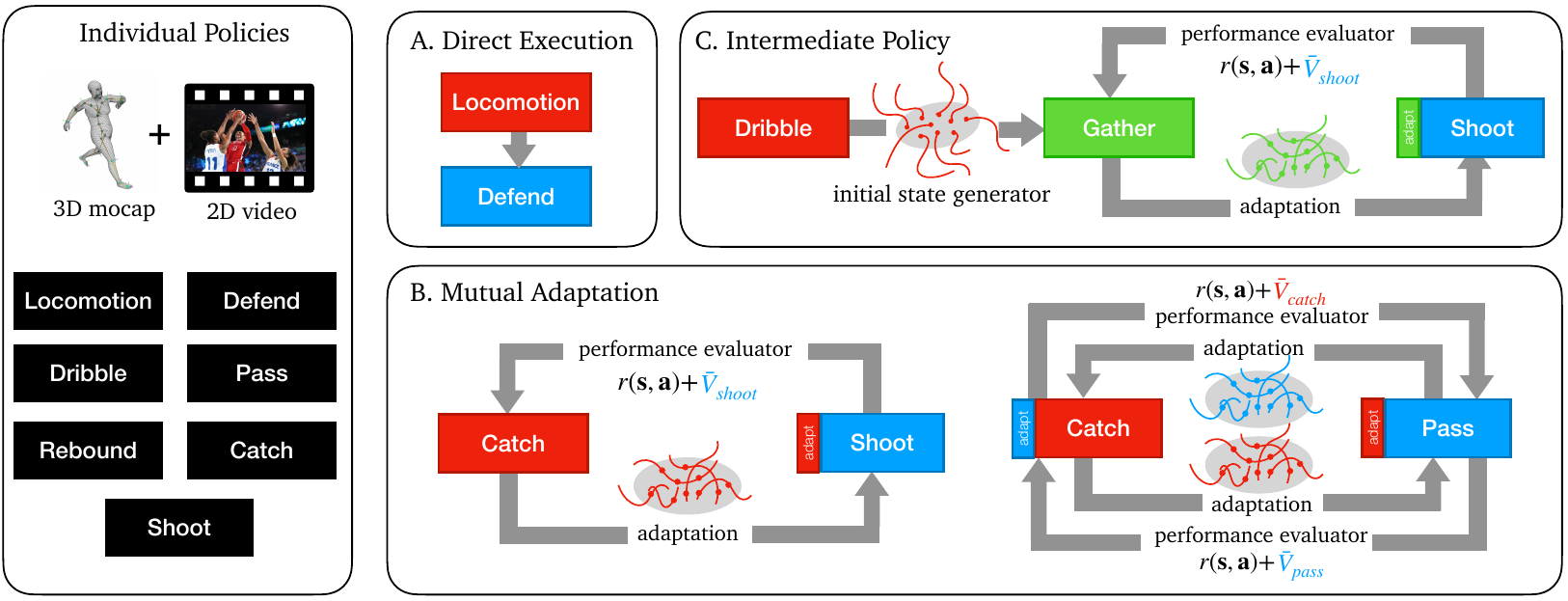}
    \caption{Three transition types are illustrated by examples. This work focuses on the most challenging case, Type C transition (upper-right), which requires an intermediate policy to facilitate the transition. To train a policy for such an ill-defined subtask (green), our method utilizes the terminal states of the preceding policy (red) to provide an initial state distribution, and the succeeding policy (blue) to provide its state value function $\bar{V}_\text{shoot}$ for reward shaping. Simultaneously, the selected states from the rollouts of the intermediate policy are used to adapt the succeeding policy. Type B transition is less challenging and can be done with only adaptation and value-function-based reward shaping, without the need to train an intermediate policy specifically. Type A transition is the least challenging one and can be done by directly executing the succeeding policy from any state of the preceding policy.}
    \label{fig:main_method}
\end{figure*}

We focus the introduction to our approach on the most challenging case (Type C transition), which requires intermediate policies.
Note that the need for intermediate policies is specific to the subtask definition. Our goal in this work is to provide solutions to three types of transitions, so the user can systematically build transitions between any arbitrary pair of subtasks. Intermediate policies are needed for Type-C transitions only. As shown in Figure~\ref{fig:transition_diagram}, most transitions (Type A and B) do not require them and can be handled by simplified variants of the same method (Figure \ref{fig:main_method}). 
We ground our exposition on the task of \emph{shooting off the dribble}, a common basketball skill. By introducing an intermediate gathering policy between the pre-trained dribbling and shooting policies, our simulated player achieves a 91.8\% shooting success rate from a wide range of challenging dribbling states, including facing away from the hoop and performing complex maneuvers such as spins, pivots, and pull-up jumpers.

The conditions for transition are also crucial to the success of a challenging long-horizon task. We introduce a high-level routing policy, trained to automate subtask transitions, eliminating the need for manually defined state conditions. Similar to controls in video games, the router allows transitions to be triggered by a user or external program, balancing smoothness with responsiveness.

Although prior work has demonstrated that imitation-based RL are effective in learning from human motions, ensuring their robustness and generality remains challenging and requires substantial demonstration data. Unfortunately, structured datasets with corresponding full-body motion, detailed finger motion, and ball trajectories remain scarce. Therefore, a secondary but key contribution of our approach is demonstrating that it is possible to learn complex basketball skills from heterogeneous datasets containing both 2D videos and 3D mocap, without requiring ball trajectories or correspondence between full-body and hand motion data.

\section{Learning from Unstructured Motions}
\label{sec:low_level}

\begin{figure*}
    \centering
    \includegraphics[width=\linewidth]{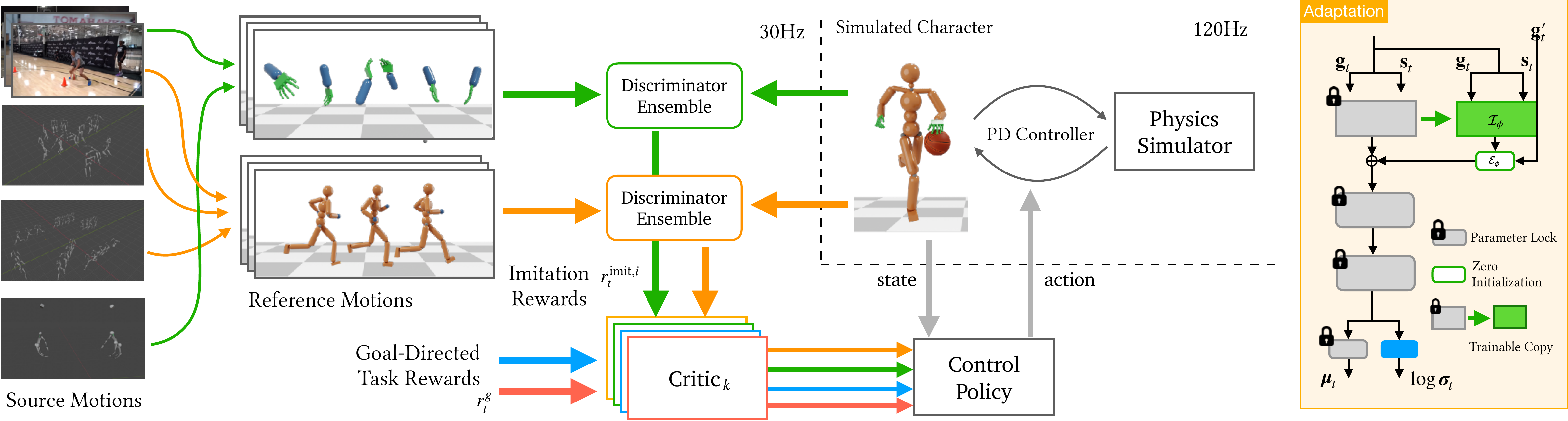}
    \caption{System architecture for primitive policy learning in our system. We decouple the full-body motions into body and hand split, and use reinforcement learning to imitate unstructured motions collected from disparate sources without needing ball trajectory reference or requiring the correspondence of the motions. From top to down on the left side, we show the screenshots of motions from online videos, body-only mocap data of basketball playing and normal locomotion, and our own captured hand motions. On the right side, we show the network structure for policy adaptation~\cite{adaptnet}. This structure allows us to introduce an optional additional goal input $\mathbf{g}^\prime_t$ during adaptation for policy transition training, and is suitable for our training strategy for pivoting foot control when transitioning to the shooting policy from dribbling motions (see the supplementary materials for the details). %
    }
    \label{fig:iccgan}
\end{figure*}

Figure~\ref{fig:iccgan} shows our system architecture for primitive policy learning from unstructured motions.
We use Proximal Policy Optimization (PPO)~\cite{schulman2017proximal} as the backbone reinforcement learning algorithm and employ an adversarial imitation learning framework~\cite{iccgan,composite} to train the primitive policies.
Specifically, for the \emph{shooting-off-the-dribbling} task, we train policies for two well-defined subtasks: dribbling in an arbitrary target direction with a randomly given speed, and shooting the ball into the hoop. Our objective is to train a character capable of dribbling and shooting on demand in real time. This goal makes it impractical to rely on tracking a fixed set of motion trajectories, as done in prior work~\cite{wang2024skillmimic}. Achieving the desired flexibility requires access to a large-scale, diverse set of reference motions, which necessitates relaxing the assumption of structured motion datasets with one-to-one correspondences between gross body motion, detailed hand motion, and basketball trajectories. To address this challenge, we:
a) incorporate multiple motion data sources, including 3D motion capture and 2D video data;
b) decouple full-body motions into several groups~\cite{composite,liu2018learning}, reducing the need for combined motions across different body parts; and
c) rely on task rewards to eliminate dependency on corresponding basketball trajectories, as collecting or generating them is highly challenging.
We refer to the supplementary materials for implementation details and hyperparameters for policy training.
Despite the strength of our approach for primitive policy learning, 
our method for policy transition (Section~\ref{sec:high_level}) does not have a special requirement on how primitive policies are trained, and can be combined with other existing, primitive skill learning methods.

\subsection{Unstructured Data Sources}
Public motion capture datasets, such as LAFAN1~\cite{harvey2020robust}, AMASS~\cite{mahmood2019amass}, the CMU Mocap Dataset~\cite{cmumocap}, provide %
full-body motion data but lack detailed hand motions, including wrist and finger dynamics. On the other hand, numerous datasets focus on detailed hand motions but do not include corresponding full-body motion 
data~\cite{fan2023arctic,wang2024dexcap,taheri2020grab, chao2021dexycb}. Video data offer the potential to capture both full-body and hand motions at scale. However, it is limited by challenges such as occlusion, depth ambiguity,  inconsistent picture quality, and motion blur, making it less reliable as a primary data source.
In this work, we use all the data sources described above, including:
\begin{itemize}
    \item Internet videos: We use ExAvatar~\cite{moon2025expressive}  and TRAM~\cite{wang2025tram} to extract hand and body poses from online videos, respectively. 
    \item Full-body-without-hand mocap data: running motions from LAFAN1~\cite{harvey2020robust} and other motions from the CMU Mocap Dataset~\cite{cmumocap} to train the base skills. %
    \item Hand-only data: We recorded our own finger motions using Rokoko Smartgloves, capturing a subject shooting in place and dribbling within a short range.
\end{itemize}

\noindent Notably, we do not assume access to the corresponding ball trajectories in our dataset. While it is technically feasible to extract ball trajectories from videos or record ball states using motion capture systems, acquiring high-quality and large-scale trajectories remains labor-intensive. Moreover, avoiding dependency on basketball-specific trajectories allows us to utilize non-basketball motion data, such as the running motion from LAFAN1 dataset.

\subsection{Learning to Dribble}
To fully take advantage of the unstructured motion datasets for training a dribbling policy, we group full-body poses into three categories: lower body, upper body, and hands. Unlike typical grouping schemes~\cite{liu2018learning,bae2023pmp}, we treat the two hands (including wrist rotations) as a separate group from the arms, with the elbows serving as the root links. This separation facilitates the use of hand-only motions.
Putting two hands into one group improves coordination between the hands and helps prevent unnatural poses, such as dribbling with both hands simultaneously.

We train the dribbling policy using reinforcement learning in a GAN-like architecture, combined with a multi-objective learning framework~\cite{composite} to balance imitation and task-specific goals. The policy is guided by two imitation objectives, one for hands and one for the reset body parts, with rewards provided by discriminators that process partially observable motions, as shown in Figure~\ref{fig:iccgan}. Additionally, two task-specific rewards are employed: one for velocity-controlled navigation and another for dribbling.
We refer to the supplementary materials for more details. During training, the policy autonomously explores physical interaction with the ball while being guided by the two task rewards and partially observable reference motions.

To adhere to the basketball rules and ensure valid interactions between the simulated character and the ball, we implement two types of violation detection as part of our goal-directed reward functions that consider: (1) invalid contact between the ball and other body parts besides the hands; and (2) traveling when the ball is held. 
The input to the dribbling policy includes the current pose of the character and of the ball, a target velocity $\mathbf{v}_\text{target}$, and a variable indicating the dribbling state of the ball.
The target velocity is generated randomly during training and given by the user via the joystick during interactive control.

\subsection{Learning to Shoot}\label{sec:shoot_policy}
For training the shooting policy, we do not consider any body part grouping but directly perform imitation learning using three full-body demonstrations of jump shooting. 
We use one full-body imitation objective and a task-specific reward measuring the shot accuracy and ball-holding performance before the shot is taken. 
We refer to the supplementary materials for
the details of state space, action space, and reward definition.

\section{Learning Intermediate Subtasks}\label{sec:middle_level}
In basketball terms, the transition between dribbling and shooting is defined by another subtask called ``gathering''. This intermediate but critical subtask not only requires the character to rapidly stop and catch the ball, but also to adjust ball-hold poses and body orientations for jump shooting while maintaining balance. However, gathering does not have clearly defined initial and terminal states, preventing simply training an imitation policy or chaining it between the dribbling and shooting policies.

To train such an intermediate subtask, our method utilizes the terminal states of rollouts generated by the preceding policy (dribbling) to provide an initial state distribution. %
Without defining a fixed set of terminal states of dribbling, the gathering policy is expected to take over the character from any dribbling state. Thereby, random states drawn from the dribbling rollouts will be taken as the initial state for the gathering policy.
Meanwhile, we utilize the state value function of the succeeding policy (shooting) to shape the reward function, so that the character learns to reach a state from which the shooting policy is likely to succeed (cf. Type C in Figure~\ref{fig:main_method}). The reward for the gathering policy is defined as:
\begin{equation}
\label{eq:rgather}
    r_\text{gather} = \begin{cases}
        -1 \qquad\qquad\qquad\quad\,\text{if any violation is detected,} \\
        r_\text{pose} + 0.25\textsc{Clip}\left(\bar{V}_\text{shoot}(\mathbf{s}_t, \mathbf{g}_t), -v, v\right) \,\text{otherwise.}
        \end{cases}        
\end{equation}
$r_\text{pose}\in[0, 1]$ is a reward term evaluating the ball holding performance, involving finger and palm poses related to the ball and body orientation related to the hoop (see the supplementary materials for details). We define $\bar{V}_\text{shoot}$ as the accumulated \emph{task} reward from state $\mathbf{s}_t$ following the shooting policy, bounded by $[-v,v]$. We employ PopArt~\cite{van2016learning} to perform value normalization on the critic~(value network) output in PPO to obtain $\bar{V}_\text{shoot}$. We set $v=1$ in all of our experiments. We refer to the supplementary materials for details of the reward function definition.
While the $\bar{V}_\text{shoot}$ term can help guide the policy to generate poses preferred by the shooting policy, the heuristic reward term $r_\text{pose}$ for gathering is still necessary to steer the character to a near-shooting pose, since the value function will generally produce low values when the character is far from shooting poses, leading to ineffective guidance.

To improve the generalizability of the shooting policy and make it better cooperate with the gathering policy, we further adapt the shooting policy, in tandem with the gathering policy training, by taking the states produced by the gathering policy as the initial state for the shooting policy, using the approach of latent space manipulation from AdaptNet~\cite{adaptnet}. 

Instead of adapting the shooting policy \emph{after} the gathering policy is trained, a key algorithm design choice is to adapt the shooting policy simultaneously \emph{during} training of the gathering policy. Specifically, we filter out the ``good'' states encountered during the training of the gathering policy and use them as the initial states for adapting the shooting policy. As the policy is adapted, its corresponding state value function is also updated and being used by the reward function of gathering policy. A good state has a score greater than $-v$ when evaluated by the evolving $\bar{V}_\text{shoot}.$ %
Since initially the poses generated by the gathering policy are distinctively different from the high-value states for the pre-trained shooting policy, we bootstrap the learning by randomly permitting $25\%$ of the states with value estimations less than $-v$, as long as the ball is held in hand and the character is approximately facing the hoop. These states are used as additional initial states to adapt the shooting policy. 

The gathering policy is trained using the same goal state as we obtain the pretrained shooting policy (Section~\ref{sec:shoot_policy}) but with an additional variable to indicate the pivoting foot to prevent traveling.
We ignore the traveling problem during the pertraining of the shooting policy as it would not happen when the policy simply imitates the shooting motions in the reference, where the subject directly jumps up for shooting. With the introduction of gathering, we extend the shooting policy during adaptation and also include the pivoting foot indicator in the goal state of the adapted shooting policy.

\begin{figure}
    \centering
    \includegraphics[width=.85\linewidth]{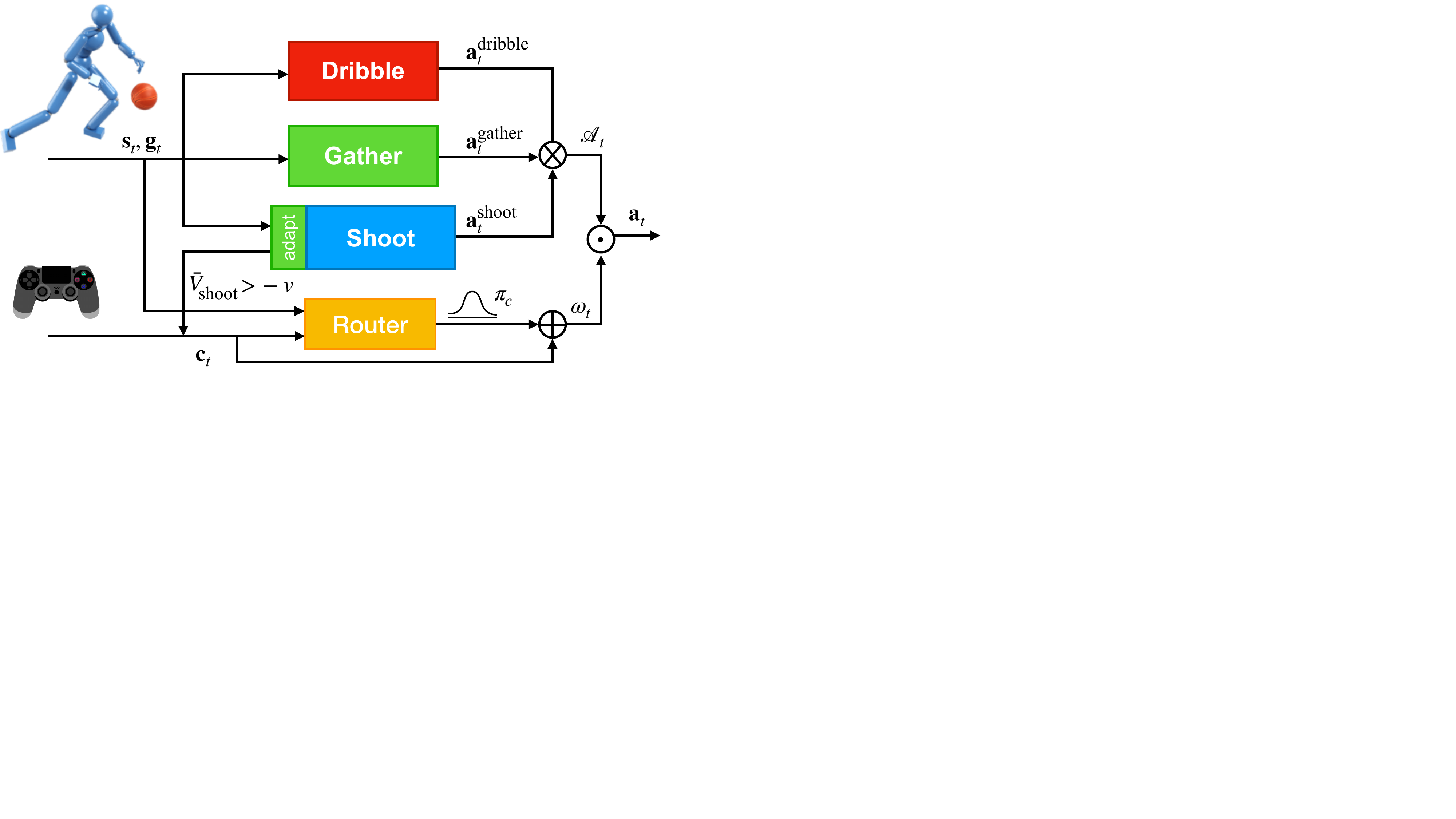}
    \caption{The high-level routing policy assembles the subtask policies to perform shoot-off-the-dribble task. It takes as input the user command, the current state, and the goal vector, and outputs the weights for linear combination of the actions from the subtask policies. $\oplus$ denotes element-wise add, $\otimes$ denotes concatenation, and $\odot$ denotes dot product. }
    \label{fig:overview_highlevel}
\end{figure}

\section{Composing Policies for Long-Horizon Tasks}\label{sec:high_level}
One possible approach to transitioning between policies is to initiate gathering when the user commands the character to take a shot, and transition to the shooting policy when the reward of the gathering policy (Eq.~\ref{eq:rgather}) is higher than the $-v$ threshold. However, using such a heuristic often results in the character stalling in a ball-holding pose without transitioning to the shooting policy or failing to catch the ball after the transition. 
To address this issue, we introduce a high-level routing policy that assembles the subtask policies to perform a long-horizon task (Figure \ref{fig:overview_highlevel}).

The primitive policies in our studied case synthesize the unstructured motions for subtask execution instead of tracking a fixed set of or a given full-body reference.
This makes motion imitation, during high-level policy training, inapplicable.
To avoid the high-level policy averaging too much on the outputs of the primitive policies and generating unnatural motions,
we model the high-level policy as a router function.
Unlike the conventional policy routing technique \cite{bacon2017option,tessler2017deep}, 
our routing policy performs soft merging instead of activating only one policy at a time step. We define a reference command $\mathbf{c}_t$, as a one-hot vector to heuristically indicate whether the character should dribble, gather, or shoot the ball. The routing policy takes $\mathbf{c}_t$ as input and outputs an offset from $\mathbf{c}_t$ to produce the weights for linear combination of the actions from the primitive policies:
\begin{equation}
    \boldsymbol{\omega}_t = \mathbf{c}_t + \pi_\text{c}(\mathbf{s}_t,\mathbf{g}_t, \mathbf{c}_t), \quad \mathbf{a}_t = \boldsymbol{\omega}_t \cdot \mathcal{A}_t.
\end{equation}
Here $\pi_\text{c}$ is the routing policy and $\mathcal{A}_t = [\mathbf{a}_t^\text{dribble}, \mathbf{a}_t^\text{gather}, \mathbf{a}_t^\text{shoot}]$ is the collection of the deterministic output from the three subtask policies, i.e. 
$\mathbf{a}_t^\text{dribble} = \mathbb{E}[\pi_\text{dribble}(\cdot | \mathbf{s}_t, \mathbf{g}_t)]$, 
$\mathbf{a}_t^\text{gather} = \mathbb{E}[\pi_\text{gather}(\cdot | \mathbf{s}_t, \mathbf{g}_t]$,
$\mathbf{a}_t^\text{shoot} = \mathbb{E}[\pi_\text{shoot}^{+}(\cdot | \mathbf{s}_t, \mathbf{g}_t)]$ and $\pi_\text{shoot}^{+}$ indicates the shooting policy after adaptation.
Note that the definition of $\mathbf{g}_t$ with respect to each subtask is different. We refer to the supplements for details.

The reference command for dribbling is $\mathbf{c}_t = [1, 0, 0]$ and becomes $[0, 1, 0]$ after receiving an external command to shoot.  When $\bar{V}_\text{shoot}(\mathbf{s}_t, \mathbf{g}_t) > -v$, we set $\mathbf{c}_t = [0, 0, 1]$, which implies that the shooting policy may be able to take over the character. 
At the beginning of training, $\pi_\text{c}$ outputs small values and the exploration will start based on the given $\mathbf{c}_t$ as guidance. As training goes on, the final weight $\boldsymbol{\omega}_t$ may diverge from $\mathbf{c}_t$ in order to generate more stable transitions between different policies.

To encourage $\pi_\text{c}$ to generate one-hot-like weights $\boldsymbol{\omega}_t$ through which only one policy dominates at each time step,
we define the training reward for $\pi_c$ as follows: 
\begin{equation}\label{eq:r_high_level}
    r = \begin{cases}
    r_\text{gather} \,\;\text{if the gathering policy dominates the control,} \\
    0.5 I_\text{switch} + r_\text{shoot}  \,\; \;\: \text{ if the shooting policy dominates,} \\
    0 \qquad\qquad\qquad\qquad\qquad\qquad\qquad\qquad\;\: \,\;\text{otherwise,}
    \end{cases}
\end{equation}
where we consider a policy dominates the control if its associated weight in $\boldsymbol{\omega}_t$ is larger than the sum of the other two. $I_\text{switch} = 1$ only for the first time when the dominant policy switches from \emph{gather} to \emph{shoot} and $I_\text{switch} = 0$ otherwise. 
The detailed definitions of $r_\text{gather}$ and $r_\text{shoot}$ are provided in the supplementary materials. 
The high-level policy after training can achieve almost 100\% success rate for ball gathering from dribbling states and an overall shot percentage of 91.8\% (see Section~\ref{sec:exp}).

To reduce the inference time consumption, after the training of the high-level composer policy, we perform an additional distillation process to compress the hierarchical policy into a single neural network. This process is done in a simple supervised learning manner with samples generated online through the physical simulator. The distilled policy during our experiments can achieve the same performance as the hierarchical policy with only trivial errors.

\section{Other Skill Transitions}
The same method used for \emph{shoot-off-the-dribble} is directly applied to \emph{pass-off-the-dribble}. Both of them are defined as Type C transitions as shown in Figure~\ref{fig:main_method}. Our system also supports a range of other skills, including rebounding, catching, defending, general locomotion, and various transitions, as summarized in Figure~\ref{fig:transition_diagram}. Type A transitions are achieved through direct execution of the succeeding policy. Type B transitions serve as a simplified version of Type C, requiring reward shaping of the preceding policy using the value function of the succeeding one, along with adaptation of the policy using rollouts generated by the preceding one (cf. Figure~\ref{fig:main_method}).

A variation of Type B transitions is employed when training catching and passing transitions (Figure~\ref{fig:main_method}, bottom-right). After individually pre-training both policies, we fine-tune them jointly through co-adaptation and co-reward shaping using two interacting agents. 
While adapted to diverse initial poses produced by the terminal states of the passing and catching policies,
this enables the passing agent to learn to make catchable throws, and let the catching agent learn to receive the ball in a state ready for an immediate pass/shot.

\section{Experiments}\label{sec:exp}

We conduct experiments to evaluate our skill composition framework for physics-based character control. While we provide qualitative results for all skill transition types in Section~\ref{sec:qualitative} (see also companion video), our quantitative analysis in Section~\ref{sec:quant} focuses on the most challenging Type C transition from dribbling to shooting that requires the use of an intermediate gathering policy, as shown in Figure~\ref{fig:main_method}. 
Our experiments utilize IsaacGym~\cite{makoviychuk2021isaac} as the underlying physics engine.
All policies are trained using PPO~\cite{schulman2017proximal}. 
Other implementation details, including policy training procedures and hyperparameters, are provided in the supplementary materials.

\begin{figure}
    \centering
    \includegraphics[width=.32\linewidth]{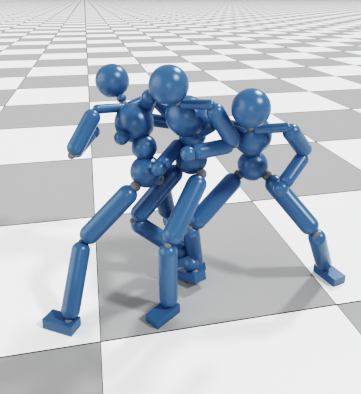}
    \includegraphics[width=.32\linewidth]{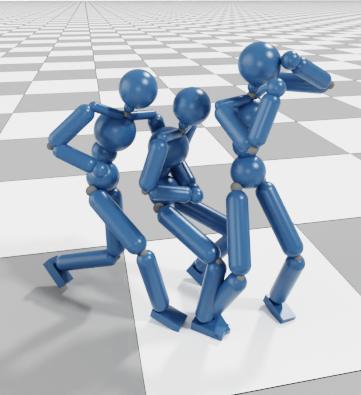}
    \includegraphics[width=.32\linewidth]{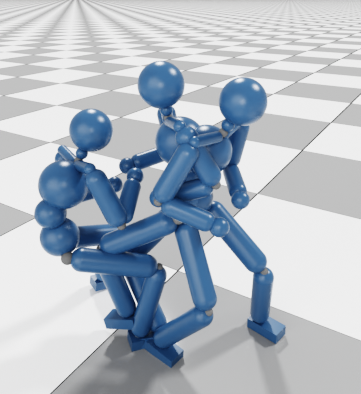}
    \caption{Examples of the gathering reference motions extracted from videos without hand. From left to right, the subject is respectively pivoting to spin clockwise with the right foot on the ground, catching the ball directly during dribbling forward for shooting, and performing a fast in-place spinning counter-clockwise with the right foot pivoting on the right foot.}
    \label{fig:gather_ref_motions}
\end{figure}

\begin{table}[t]
    \caption{Reference motions for primitive policy training. All videos are collected online. The \textbf{Dribble} mocap (body) data is collected from CMU Mocap dataset.
    \textbf{Locomotion} data are running motions from LAFAN1.
    All mocap (hand) motions are collected by ourselves.
    Additionally, we take 3 demonstrations of jump-shooting motions (full body with hands) to train the shooting policy.
    We mixedly use gathering and catching motions for catching and rebounding. All motions do not include ball trajectories.}
    \label{tab:motions}
    \centering
    \begin{tabular}{rc|rc}
    \toprule
    \multicolumn{1}{l}{\textbf{Source}} & \textbf{Length} & \multicolumn{1}{l}{\textbf{Source}} & \textbf{Length} \\
    \midrule

    \multicolumn{1}{l}{\textbf{Dribble}} & \textit{(body/hand)}   & \multicolumn{1}{l}{\textbf{Defensive Stance}} & \textit{(body/hand)} \\
       video & 48.8s/36.1s & video & 9.3s/1.4s \\
       mocap & 58.2s/30.5s & mocap & 25.1s/25.1s \\
     \multicolumn{1}{l}{\textbf{Gather}} & \textit{(body/hand)} & \multicolumn{1}{l}{\textbf{Locomotion}} & \textit{(body)} \\
     video & 30.3s/7.7s & mocap & 141.4s\\
     \multicolumn{1}{l}{\textbf{Catch}} & \textit{(body/hand)} & \multicolumn{1}{l}{\textbf{Pass}} & \textit{(full-body)} \\
     video & 10.1s/10.1s & video & 4.5s\\
    \bottomrule
    \end{tabular}
\end{table}

\subsection{Data Preprocessing}
As detailed in Section~\ref{sec:low_level}, we use ExAvatar\cite{moon2025expressive} and TRAM \cite{wang2025tram} to extract 3D hand and body poses from videos and perform motion synthesis via imitation learning for primitive policy training. The video data is collected exclusively from publicly available sources.
Additionally, we enrich the dataset by combining the processed motions with publicly available mocap data and a small set of hand motions that we captured ourselves.

Table~\ref{tab:motions} summarizes the data sources and the motion lengths in our reference motion dataset. Most of the \textbf{Dribble} motions extracted from videos consist of in-place movements due to the limitation of the pose estimation models when facing fast movement subjects. We incorporate normal running motions from LAFAN1~\cite{harvey2020robust} to enable the policy to learn locomotion with fast movements and directional changes during dribbling. \textbf{Gather} motions
are extracted from 13 video demonstrations (see examples in Figure~\ref{fig:gather_ref_motions}). Due to occlusion and rapid hand movements, we obtained only a limited set of ball-holding hand motions during the gathering phase.
The \textbf{Defensive Stance} motions include stance poses while the subject stretches the arms to perform blocking or screening.
For shooting, we specially use three full-body demonstrations of jumping shooting, including hand poses.
Our dataset does not include ball trajectories. Instead, the control policy explores valid interactions with the ball on its own under the guidance of the task rewards.

\subsection{Quantitative Evaluation}\label{sec:quant}
To evaluate the quality of Type C transitions generated by our approach, we analyze the performance of the shoot-off-the-dribble task by focusing on two primary metrics: 
\begin{itemize}
    \item {\em Ball Catching Rate:} This metric checks whether the ball is successfully gathered and held by the character after a shooting command is issued and before the ball is shot.
    \item {\em Shot Percentage:} This measures the ratio of successful field goals to attempted field goals. Attempts that fail due to unsuccessful ball catching or a lack of response to the shooting command are also included in this metric.
\end{itemize}
We compare our method against three baselines which offer alternative approaches for policy composition. These baselines represent \emph{zero-shot} and \emph{sequential} approaches that have been explored in different application domains for skill chaining. 
\begin{itemize}
    \item {\em\textbf{DirectExecution} (Dribble$\tiny\rightarrow$Pretrained Shoot):} This baseline uses the pretrained shooting policy ($\pi_\text{shoot}$) without any adaptation or intermediate gathering policy. 
    We use grid search on the state value of the shooting policy to find out the optimal transition, which is similar to the implementation in prior work~\cite{iccgan}.
    
    \item {\em\textbf{NoAdapt} (Dribble$\tiny\rightarrow$Gather$\tiny\rightarrow$Pretrained Shoot):} This approach incorporates a gathering policy but no adaptation is applied to $\pi_\text{shoot}$. The function of $\bar{V}_\text{shoot}$, thereby, is fixed during gathering policy training. Like {\em DirectExecution}, this is zero-shot with the state value used for policy switching. 
    
    \item {\em\textbf{SequentialChaining} (Dribble$\tiny\rightarrow$Shoot) :} The sequential approach
    leverages terminal states from the preceding policy to train the succeeding policy and improve transition success~\cite{clegg2018learning, liu2017learning, wang2024skillmimic}. A shooting policy is trained from random initial states provided by the dribbling policy.
\end{itemize}
Additional comparisons to other policy-switch strategies and detailed ablation studies are provided in Section~\ref{sec:ablation}.

\begin{figure}
    \centering
    \includegraphics[width=.32\linewidth]{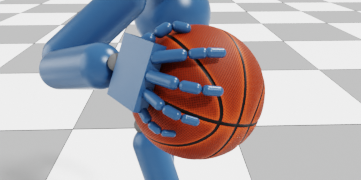}
    \includegraphics[width=.32\linewidth]{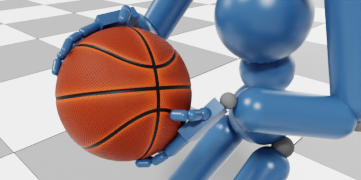}
    \includegraphics[width=.32\linewidth]{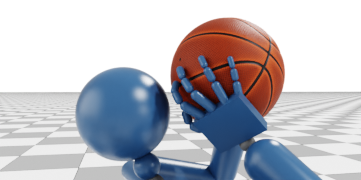}
    \caption{Learned ball-hand interactions by policies trained with our approach. From left to right, the snapshots are captured during the character dribbling, gathering and shooting the ball, respectively.}
    \label{fig:hand_ball_interactions}
\end{figure}

\begin{figure}
    \centering
    \includegraphics[width=\linewidth]{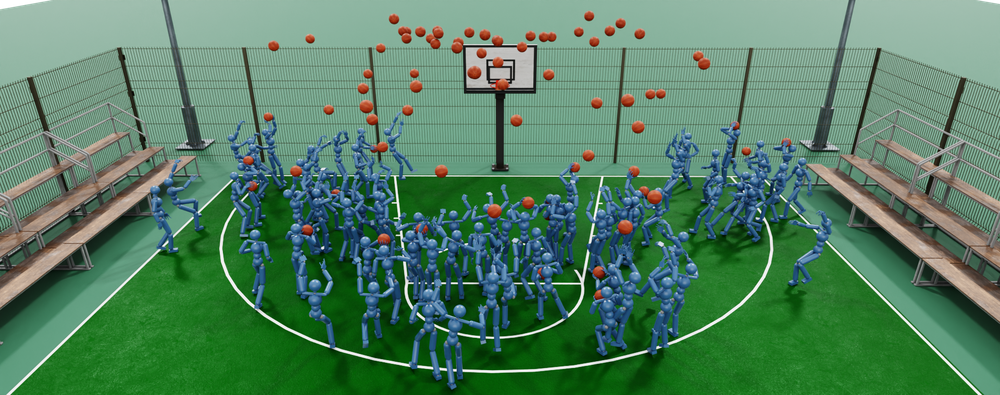}
    \caption{The testing scenario to count the shot percentage.}
    \label{fig:shoot}
\end{figure}

\subsubsection{Ball Catching Rate}
\begin{figure}
\centering 
\includegraphics[width=\linewidth]{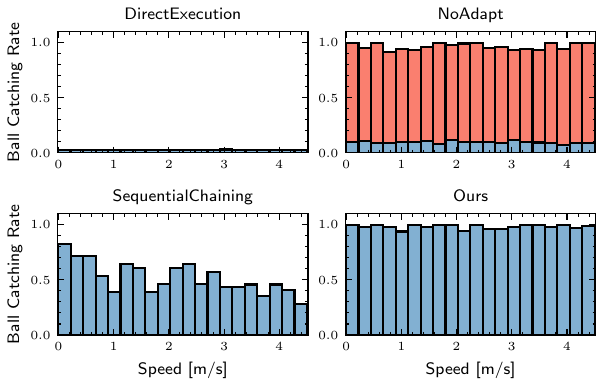} 
\caption{Ball catching rates as a function of dribbling speed. The \textbf{DirectExecution} baseline fails almost entirely due to the incompatibility between policies. The \textbf{NoAdapt} baseline shows high success rates for the gathering policy (red bars), but suffers from poor generalization of the shooting policy after the shooting policy takes over the character (blue bars). The \textbf{SequentialChaining} baseline performs better overall but underperforms compared to the dedicated gathering policy in the \textbf{NoAdapt} method. Our method achieves consistently high ball catching rates across speeds.
} 
\label{fig:gather_rate} 
\end{figure}

Figure~\ref{fig:gather_rate} shows the ball catching rate for different methods as a function of the character's dribbling speed at the time a shooting command is issued. Our method shows robust performance across all speeds, maintaining a high ball catching rate, while the baselines exhibit significant limitations.

The \textbf{Direct Execution} baseline fails almost entirely, with a $0.7\%$ catching rate across all speeds. This result highlights the incompatibility between the dynamic dribbling state and the unadapted shooting policy, consistent with prior findings~\cite{iccgan} that direct policy switching requires the preceding policy to transition the system into a compatible state for the succeeding policy.
For the \textbf{NoAdapt} baseline, the red bars in Figure~\ref{fig:gather_rate} indicate that the gathering policy performs well at catching the ball, achieving a high 
average catching rate across all speeds. However, the pretrained shooting policy struggles to generalize, resulting in significant performance drop.
Even when adapting the shooting policy during the training of the gathering policy (as shown in our supplementary materials), the average ball catching rate remains low unless the state value is explicitly incorporated as a reward term for the gathering policy (cf. Eq.~\ref{eq:rgather}). 
This result underscores the difficulty of achieving effective cooperation between the gathering and shooting policies through random exploration alone during training. 
The \textbf{SequentialChaining} baseline, which trains a single policy for both gathering and shooting, achieves a higher average ball catching rate %
than the \textbf{DirectExecution} baseline but underperforms compared to the 
 \textbf{NoAdapt} that utilizes a separate 
gathering policy.  
This degradation suggests that the lack of explicit phase division between gathering and shooting hinders policy effectiveness. %

In contrast, our method achieves a $98.3\%$ ball catching rate across all dribbling speeds, as shown in the bottom-right plot of Figure~\ref{fig:gather_rate}. By dividing the task into distinct phases and leveraging state value as a reward signal, our framework ensures smooth cooperation between policies, enabling reliable ball gathering and shooting even under highly dynamic conditions. These results underscore the value of structured phase division and reward shaping in multi-phase tasks.

\subsubsection{Shot Percentage}

\begin{table}
    \caption{Shot percentage of our approach when the character dribbles at different approaching directions toward the hoop. ``Orth.'' stands for the orthogonal direction. Each of the reported numbers is obtained by considering an angle range of $\pi/2$. The overall shot percentage is 91.8\%. We refer to the supplementary material for visualized results.}
    \centering
    \begin{tabular}{l|cccc}
    \toprule
    & Facing & Opposite & Left Orth. & Right Orth. \\
    \midrule
       Shot Percentage  & 95.4\% & 90.4\% & 92.7\% & 92.4\%\\
    \bottomrule
    \end{tabular}
    \label{tab:shot_percent}
\end{table}

\begin{figure*}
    \centering
    \includegraphics[width=\linewidth]{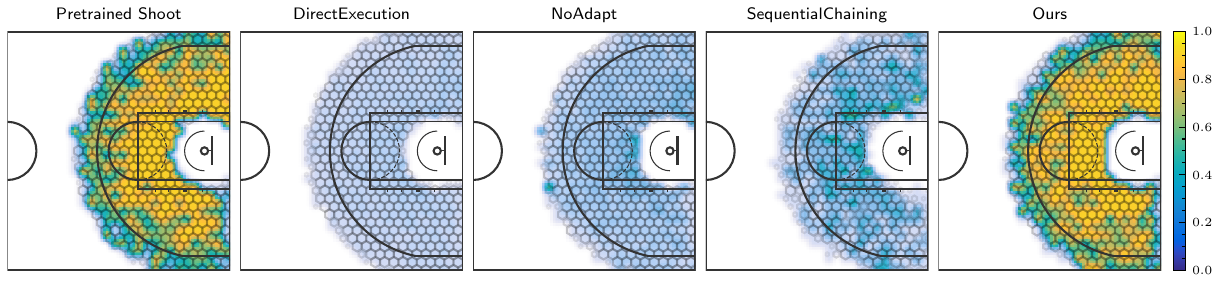}
    \caption{Heatmap of shot percentages for the baselines and our method across different positions and approaching directions on the court. The first subplot shows the performance of the vanilla pretrained shooting policy, where the character and ball states are initialized using reference motions rather than random dribbling states, achieving an overall shot percentage of 93.0\%. For baselines, directly applying the pretrained shooting policy or using combined gathering and shooting strategies leads to poor performance (1.3\%, 6.1\%, and 12.7\% overall shot percentages, respectively), as they fail to handle the dynamic and diverse states arising from dribbling. Our method achieves a shot accuracy of 91.8\%
    nearly matching the vanilla pretrained policy,
    demonstrating its ability to adjust character poses dynamically during the gathering phase and perform accurate shots even in challenging scenarios.}
    \label{fig:shot_percent_baselines}
\end{figure*}

We evaluate the shot percentage with the character under varying positions and approaching directions toward the hoop.
Similar to the training setup, we define a valid shooting area as a ring between 2.5m and 7.5m from the hoop, as shown in Figure~\ref{fig:shot_percent_baselines}. This area is divided into a grid, where each grid cell has a radius of 0.5m. In each cell, we conduct 400 trials with the character dribbling at different speeds and approaching directions. 
While our approach has an overall shot percentage of 91.8\%,
\textbf{SequentialChaining} achieves a slightly higher shot percentage (12.7\%) than the other two baselines, but still performs much poorly compared to our method.
According to Figure~\ref{fig:gather_rate},
our method has a high success rate of ball catching.
Occasional failures may occur due to excessive movement speed or an extremely unstable state (e.g., fast spinning).
For shooting off the dribbling, most missed shots originate from positions that are too far from the hoop, as shown in the rightmost subplot in Figure~\ref{fig:shot_percent_baselines}, instead of the policy transition.

Additionally, Figure~\ref{fig:shot_percent_baselines} includes the vanilla performance of the pretrained shooting policy, where the character and ball states are initialized using shooting reference motions rather than random dribbling states. Despite being trained with only three demonstrations of shooting motions, the vanilla pretrained shooting policy achieves a high overall shot percentage 
in a wide area around the hoop. This success highlights its effectiveness when starting from well-aligned and familiar states. However, when applied to unseen character and ball states resulting from dribbling or gathering, the policy's performance deteriorates significantly.

Table~\ref{tab:shot_percent} further analyzes the shooting performance of our approach based on the character's approaching direction, demonstrating its adaptability across a wide range of scenarios.
As can be seen, %
the shot percentage when the character approaches the hoop in a facing direction is better than the vanilla pretrained policy.
Overall, our solution enables the policy to handle diverse and dynamic states effectively, allowing the character to adjust its pose during gathering. This includes spinning, pivoting, or turning to align with the hoop, even when the character initially faces the opposite direction to the hoop. 

\subsection{Qualitative Analysis}
\label{sec:qualitative}

\begin{figure}
    \centering
    \includegraphics[width=\linewidth]{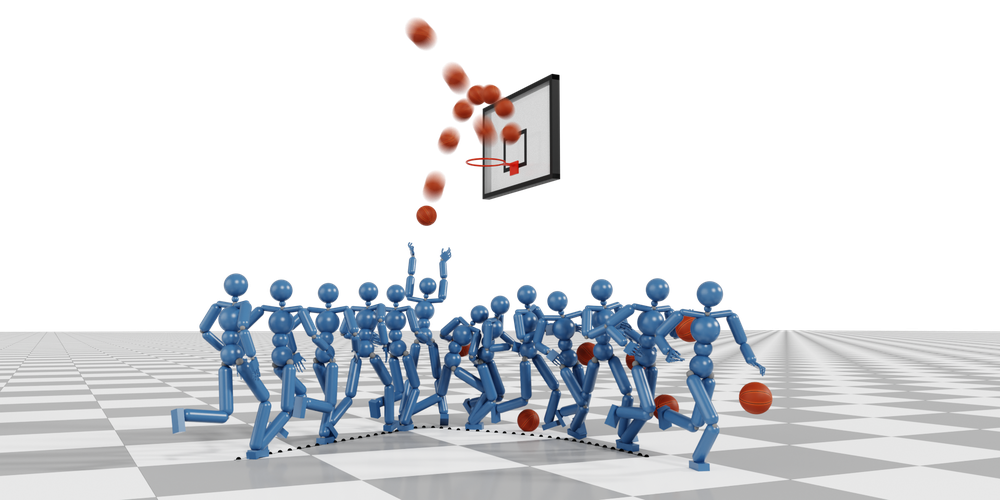}
    \caption{Transitions from locomotion to rebounding and to dribbling. The former transition is obtained via the common stance poses without additional learning (Type A transition). The latter one is obtained by adapting the pretrained dribbling policy using terminal (ball-holding) poses from a catching policy (Type B transition).}
    \label{fig:demos_rebound_dribble}
\end{figure}

Despite the absence of ball trajectory data and the reliance on normal running motions to learn locomotion during dribbling, the dribbling policy successfully synthesizes unstructured motions from multiple sources. It enables the character to perform complex tasks such as sharp turns and abrupt stops while maintaining effective ball control for arbitrary target velocities in the range of 0 to 5$\,$m/s. 
We refer to the supplementary materials for the demonstrations along with those of the other primitive skills trained using adversarial imitation learning. In all these cases, except for the shooting and passing skills, we decouple full-body motions, enabling training on unstructured, partially observable motions collected from disparate sources without needing to track any reference ball trajectories.

\begin{figure}
    \centering
    \includegraphics[width=\linewidth]{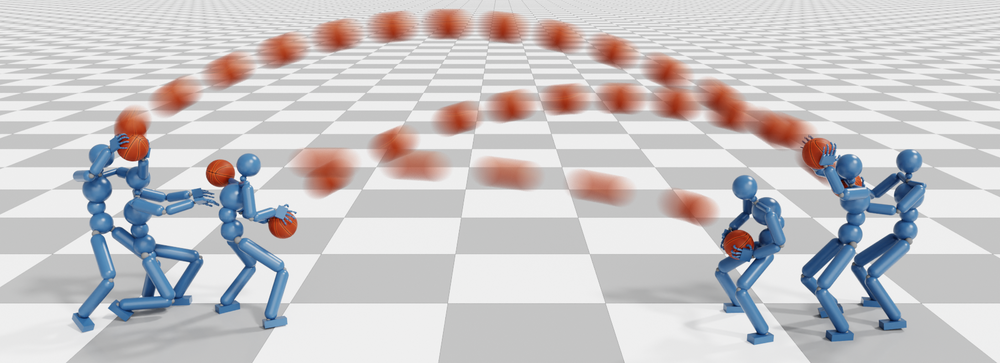}
    \caption{Passing and catching transition between different agents. The two policies are adapted using the other's state value function simultaneously.}
    \label{fig:catch_pass}
\end{figure}

\begin{figure}
    \centering
    \includegraphics[width=\linewidth]{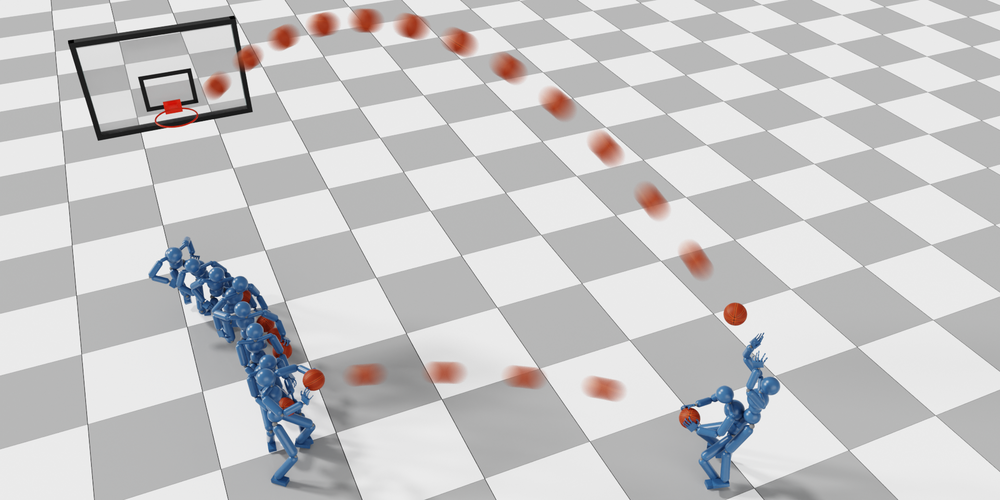}
    \caption{Cooperative behaviors between two characters. 
    Both of the passing-off-the-dribbling and shooting-off-the-catching behaviors are achieved by adapting and composing the corresponding primitive policies using our presented pipeline.}
    \label{fig:demos_catch_shoot}
\end{figure}

\begin{figure}
    \centering
    \vspace{-1em}
    \includegraphics[width=.49\linewidth]{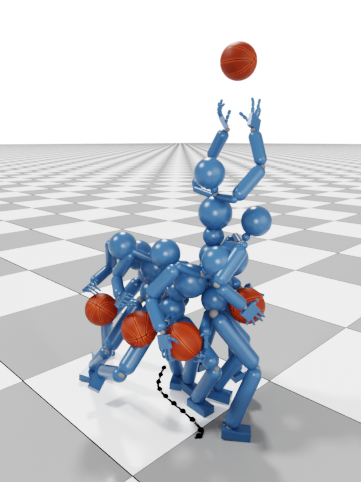}
    \hfill
    \includegraphics[width=.49\linewidth]{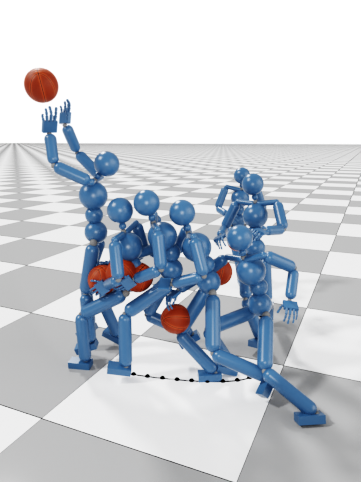}

    \vspace{-0.9em}
    \includegraphics[width=.156\linewidth]{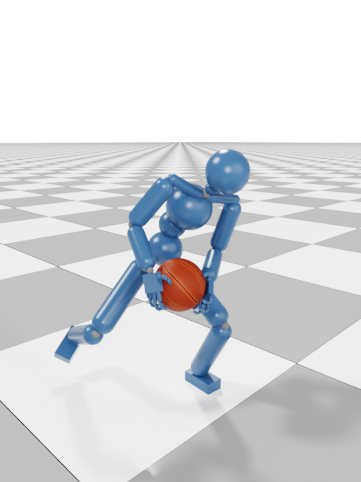}
    \includegraphics[width=.156\linewidth]{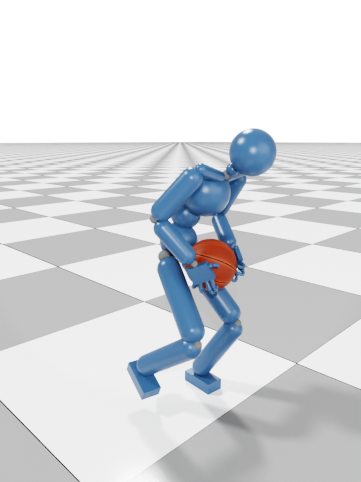}
    \includegraphics[width=.156\linewidth]{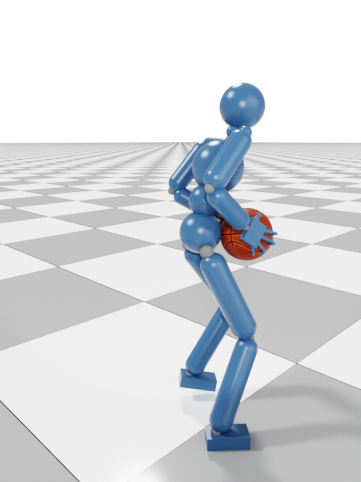}
    \hfill
    \includegraphics[width=.156\linewidth]{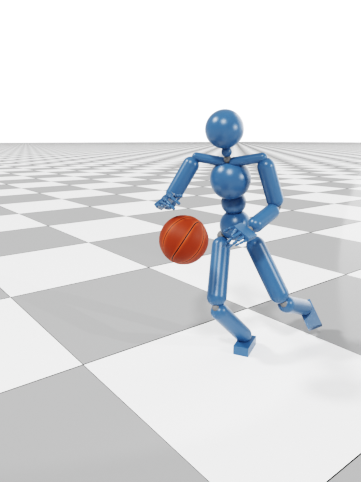}
    \includegraphics[width=.156\linewidth]{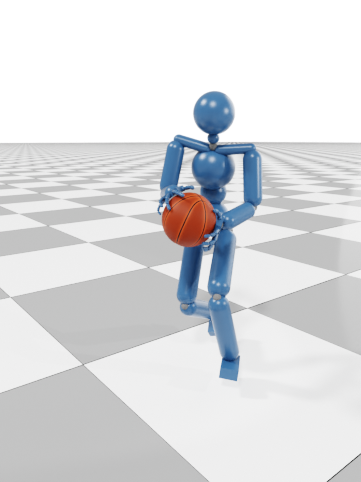}
    \includegraphics[width=.156\linewidth]{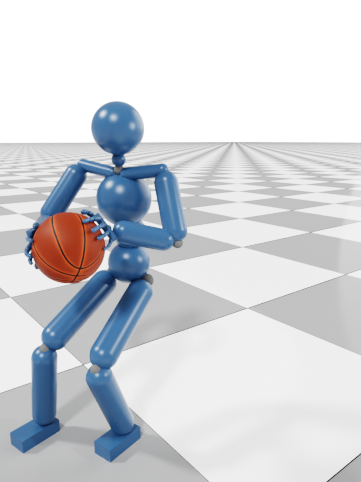}

    \vspace{-1em}
    \includegraphics[width=.49\linewidth]{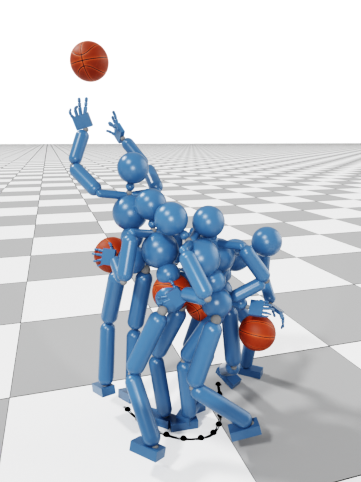}
    \hfill
    \includegraphics[width=.49\linewidth]{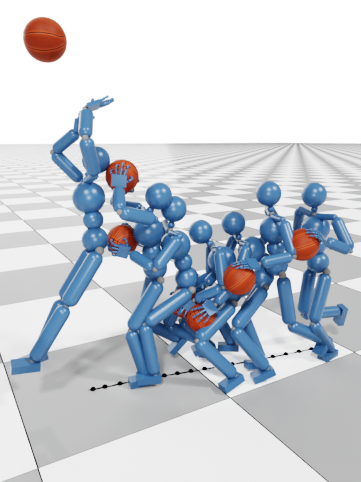}
    
    \vspace{-0.9em}
    \includegraphics[width=.156\linewidth]{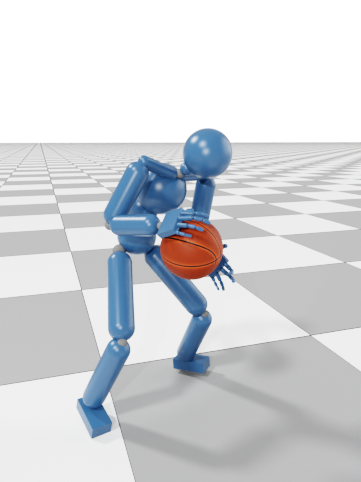}
    \includegraphics[width=.156\linewidth]{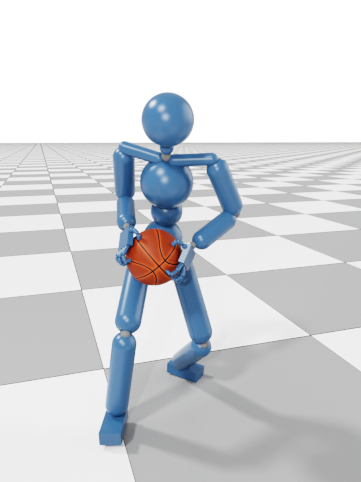}
    \includegraphics[width=.156\linewidth]{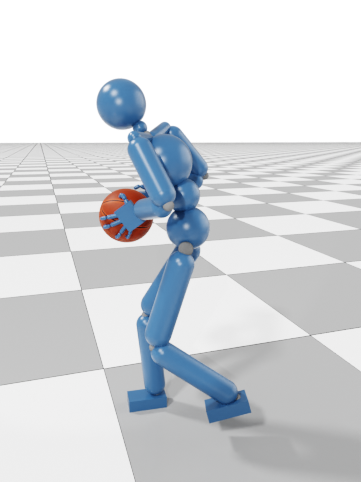}
    \hfill
    \includegraphics[width=.156\linewidth]{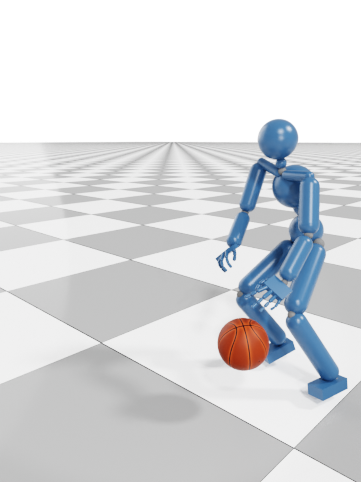}
    \includegraphics[width=.156\linewidth]{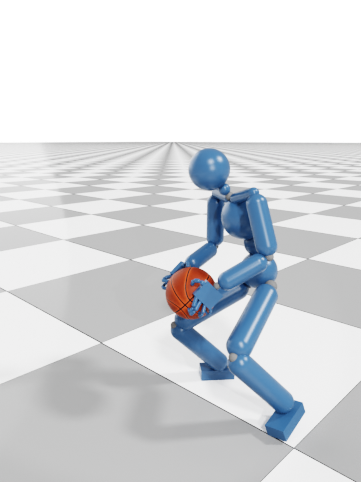}
    \includegraphics[width=.156\linewidth]{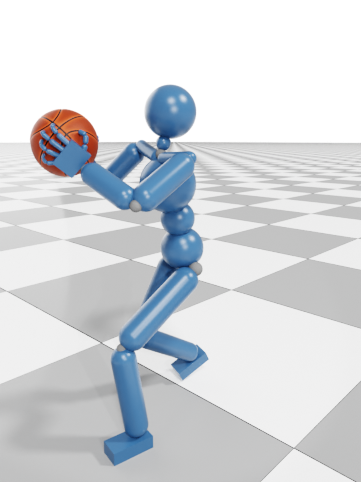}

    \caption{Demonstrations of our approach that controls the character to shoot off the dribble at arbitrary dribbling states. The snapshots under each subfigure show the keyframes where the character gathers the ball and adjusts its pose for shooting.}
    \label{fig:shoot_off_the_dribble}
\end{figure}

\begin{figure}
    \centering
    \includegraphics[width=\linewidth]{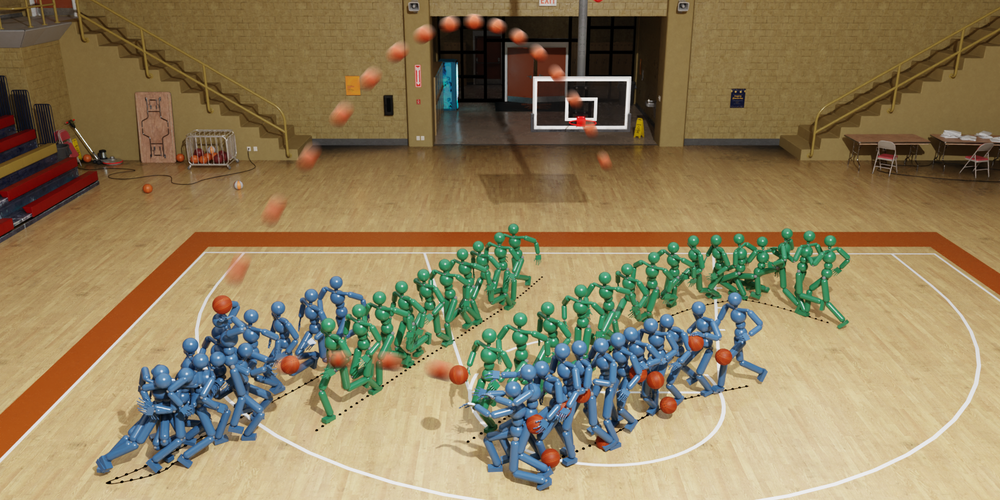}
    \caption{Four characters are controlled by human players through our control policies to perform a 2-on-2 competition in real-time.}
    \label{fig:two_interact}
\end{figure}

Figure~\ref{fig:shoot_off_the_dribble} illustrates the transitions from various dribbling states to shooting. The character is able to adjust dynamically based on context: catching the ball swiftly when possible, maintaining a dribble-like state during sharp turns before gathering the ball, or pivoting by spinning around one foot to align with the hoop. 
As shown in Figure~\ref{fig:gather_ref_motions}, the ball-gathering motions are notably different from those used for dribbling or shooting. Mixing gathering reference motions with shooting motions during policy training leads to undesirable behaviors, such as directly throwing or punching the ball toward the hoop instead of shooting (see the supplementary video).

Unlike shooting off the dribble, %
the transitions shown in Figure~\ref{fig:demos_rebound_dribble} are obtained without training an intermediate policy. 
Here, the rebounding policy is executed directly from the locomotion policy. The same applies to the transition from rebounding to dribbling, though the succeeding dribbling policy is adapted with a ball-catching policy.
Similar adaptation is applied to improve cooperative behaviors between two interacting characters by mutual adaptation, such as passing and catching the ball, and passing-off-the-dribbling and catch-and-shoot as shown in Figures~\ref{fig:catch_pass} and~\ref{fig:demos_catch_shoot}. Our system also supports competitive interactions between two characters by   
training a defending policy that enables strategic movements and defensive actions such as screening and blocking by raising hands up. Figures~\ref{fig:teaser} and~\ref{fig:two_interact} showcase interactions in which the offender players (blue) shoot while the defenders (green) block.

\section{Ablation Studies}\label{sec:ablation}
We focus our ablation studies on the evaluation of
(1)
the training strategy to obtain a good intermediate policy that bridges the gap between a preceding policy (dribbling) and a succeeding policy (shooting), and
(2)
the effectiveness of the high-level routing policy for primitive policy composition.

\begin{figure}
    \centering
    \includegraphics[width=\linewidth]{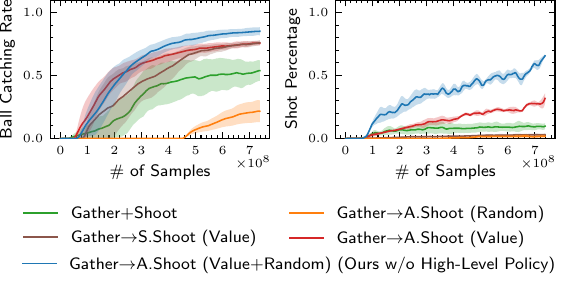}
    \caption{Learning Performance with different intermediate policy training strategies. ``S.'' stands for ``Scratch'' and means a shooting policy trained from scratch with the gathering policy. ``A.'' stands for ``Adapted'' and means a pretrained shooting policy adapted with the gathering policy. ``Value'' and ``Random'' indicate the way to transfer the character and ball states produced by the gathering policy to train or adapt the shooting policy. ``Value'' means to transfer based on the state value evaluation from the shooting policy and ``Random'' means to randomly transfer states from gathering to the shooting policy. In the case of ``Random'', the gathering policy serves only as a sample generator for the shooting policy training or adaptation, and the state-value term in Eq.~\ref{eq:rgather} thus will be ignored. The shaded ranges show the performance over three training trials.}
    \label{fig:g2s_strategy}
\end{figure}

\subsection{Intermediate Policy Training}
In Figure~\ref{fig:g2s_strategy}, we compare the learning performance of different strategies for intermediate gathering policy training and shooting policy fine-tuning.
Gather$+$Shoot is the \textbf{SequentialChaining} baseline, as we discussed in Section~\ref{sec:exp}, which treats ball gathering and shooting as a whole task in one phase.
Only a single policy is trained for that baseline, using a mixture of the gathering and shooting reference motions.
All the gathering policies shown in the figure are trained to utilize the same pretrained dribbling policy as an initial state generator,
and all the adapted shooting policies (A.Shoot) perform adaptation based on the same pretrained shooting policy.
The state-value based term in the reward function $r_\text{gather}$ (cf. Eq.~\ref{eq:rgather}) will be ignored in the baselines without the ``Value'' label.
Similar to results shown in Section~\ref{sec:exp},
the ball-catching rate is evaluated not only for the gathering policy, but also for the shooting policy to see if the transfer from gathering to shooting is stable.
To draw a fair comparison,
here we show the performance of our system without introducing the high-level policy for policy composition.

As shown in the figure,
Gather$\rightarrow$A.Shoot (Random) performs the worst among the baselines, with only a small improvement in ball catching rate at the end of training.
We also test an additional baseline (not shown in the figure) using a random state transferring strategy for the shooting policy trained from scratch, i.e. Gather$\rightarrow$S.Shoot (Random).
It performs even worse without any performance improvement throughout the training.
In contrast,
the ``Value'' baselines show a much more stable performance of ball catching rate.
It means that the introduction of the state-value reward term can effectively help the gathering policy learn how to control the character reaching a state manageable by the shooting policy.
Comparing the performance of Gather$\rightarrow$S.Shoot and that of Gather$\rightarrow$A.Shoot,
while the ball-catching rate is similar when state-value based transfer is adopted,
policy fine-tuning based on a pretrained shooting policy (A.Shoot cases) shows better performance on shot percentage.

The state-value evaluation from the shooting policy can effectively improve the performance of the gathering policy.
However,
the state-value estimator (the value network from the shooting policy under training for S.Shoot baselines or adaptation for A.Shoot baselines) is trained in tandem with the gathering policy, and is not always reliable, as it may be stuck at local optimal and cannot effectively evaluate unseen states or potentially good states (the states that are acceptable to make a field goal but the shooting policy has to be further trained or adapted to work with the given states).
Therefore, to increase the robustness of the whole system, 
our method takes a random sampling strategy complementing the state-value based transfer strategy for shooting policy adaptation, as described in Section~\ref{sec:middle_level}.

Gather$+$Shoot shows worse performance compared to the baselines of Gather$\rightarrow$A.Shoot (Value),
which demonstrates the necessity to explicitly set multiple phases for transitions between drastically different subtasks.
Using the gathering and shooting reference motions simultaneously could lead to unnatural behaviors, where the character, for example, throws out the ball while back facing the hoop or directly punches the ball out toward the hoop.
We refer to the supplementary video for the animated results.

\subsection{High-Level Routing Policy}
\begin{figure}
    \centering
    \includegraphics[width=\linewidth]{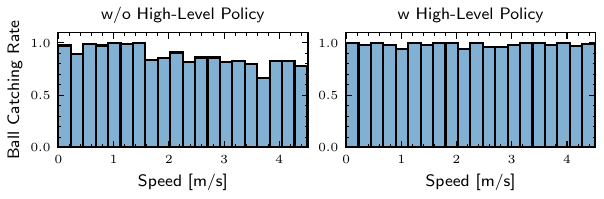}
    \caption{Ball catching rate of our system with and without the high-level policy while the character dribbles at different velocities. The overall ball catching rate is 86.0\% when no high-level policy is introduced, and 98.3\% with the high-level policy.}
    \label{fig:gather_rate_middle_vs_high}
\end{figure}

\begin{figure}
    \centering
    \qquad
    \includegraphics[width=.9\linewidth]{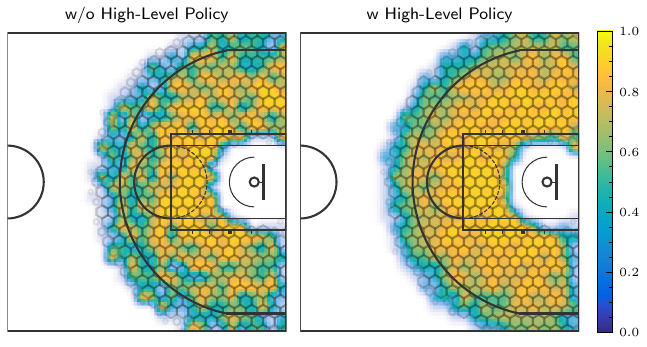}
    \caption{Heatmap of the shot percentage of our system with and without the high-level policy. The overall shot percentage is 67.4\% when no high-level policy is introduced, and 91.8\% with the high-level policy.}
    \label{fig:shot_percent_midlle_vs_high}
\end{figure}

We compare the performance of our system with and without the high-level policy in Figure~\ref{fig:gather_rate_middle_vs_high} and~\ref{fig:shot_percent_midlle_vs_high}.
Similar to other baseline comparisons, when no high-level policy is introduced,
we perform policy switches between dribbling, gathering, and shooting in a heuristic way:
the gathering policy will be called right after receiving a shooting command, and the shooting policy will take over the character if the state value evaluation based on the current state of the character and ball is higher than a threshold value.
The optimal threshold value is found by a grid search in the range $[-1, 0]$ with an interval of $0.1$.
As shown in the figures,
although both methods rely on the same primitive policies,
the high-level policy achieves better performance in terms of both the ball catching rate and shot percentage,
while also automating the transfer between those primitive policies.
Visually,
without the high-level policy,
heuristic rule-based policy switch would cause undesired behaviors like a delayed response from gathering to shooting, sticking to a ball holding pose without performing shooting anymore, or even the character falling down.
We refer to the supplementary video for the animated results.

\begin{figure}
    \centering
    \includegraphics[width=\linewidth]{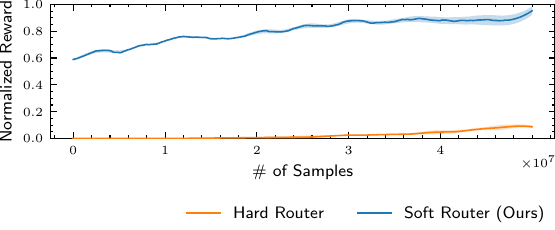}
    \caption{Learning performance when modeling the high-level policy as a soft router (ours) and that modeling the high-level policy as a hard router.}
    \label{fig:router}
\end{figure}

The high-level policy acts in a mixture-of-the-expert way.
To preserve the naturalness of the generated motions and prevent excessive averaging across the output of primitive policies,
the high-level policy trained to encourage outputting actions dominated by one primitive policy (cf. Eq.~\ref{eq:r_high_level}).
This is a ``soft routing'' strategy, as it still allows for the high-level policy to slightly merge behaviors from multiple primitive policies.
By using the soft routing strategy, 
we can easily incorporate the reference command to the output of the high-level policy.
Training of the high-level policy, thereby, starts with exploration around the reference command generated heuristically in the same way as the baseline shown in Figures~\ref{fig:gather_rate_middle_vs_high} and~\ref{fig:shot_percent_midlle_vs_high}.
This improves the training efficiency of the high-level policy largely.
In Figure~\ref{fig:router}, we compare the training performance of our soft routing strategy with the hard routing strategy.
In the hard routing case, the output of the high-level policy is modeled as a discrete one-hot vector through a softmax operation.
Without the guidance from the reference command,
the high-level policy in ``hard routing'' cases has to rely on itself to explore how to compose multiple primitive policies.
As shown in the figure,
the performance of the hard router is much lower than our soft router, and even worse than the baseline using only heuristic rules (the reference command) without the high-level policy.
This highlights that, despite having pretrained primitive policies, learning to effectively compose them remains a non-trivial challenge.
Our soft-routing approach provides a more effective way to compose policies in a flexible way.

\section{Conclusion}

Our results demonstrate the efficacy of a policy integration framework in tackling multi-phase, long-horizon tasks. By treating preceding policies as initial state generators and succeeding policies as evaluators via state value functions, we bridge the gap between drastically different subtasks, enabling seamless transitions in scenarios where intermediate states defy standard definition. This framework is flexible and works in less challenging cases without the need for a newly learned intermediate policy, or in scenarios involving sequential cooperation between multiple agents. 
The success of our approach in synthesizing diverse motion behaviors—using unstructured data sources and limited reference motions—highlights its generality, with implications beyond basketball. 
Our method can adapt to characters with different morphologies and skill stats by using styled reference motions and stats constraints during primitive policy training. Once a characteristic-conditioned policy is trained, the transitioning procedure remains the same.
The whole policy transition system is agnostic to the task of each primitive policy and can be easily extended by introducing more primitive policies.

Despite the strengths of our approach, this work has certain limitations. The motion quality falls short of the level demonstrated by skilled human players, primarily due to limited reference data. For instance, although ball-hand interactions appear plausible, the character often adopts a high dribbling posture, with hands near chest level—resembling arm positions during regular running. While ball control is reliable and rarely fails, it lacks the fluidity typical of human dribbling. Likewise, because we use only a single clip of a low jumping motion, the character exhibits minimal vertical lift when attempting to rebound the ball.

A potential future work is incorporating a biomechanically accurate hand model, which could encourage more natural physical interactions with the basketball. 
Another promising direction for future research involves using large language models to plan high-level tactical plans for multiple simulated players. 
This would allow the training of multiple autonomous agents in cooperative and adversarial settings.

\begin{acks}
This work was supported in part by the Wu-Tsai Human Performance Alliances, Stanford Institute for Human-Centered Artificial Intelligence, and Roblox. We thank Joe Gibbs Human Performance Institute and Rokoko for providing mocap data for testing.
\end{acks}

\bibliographystyle{ACM-Reference-Format}
\bibliography{reference}

\appendix
\renewcommand{\thefigure}{S\arabic{figure}}
\renewcommand{\thetable}{S\arabic{table}}
\def\theequation{S\arabic{equation}}
\renewcommand{\thealgocf}{S\arabic{algocf}}

\setcounter{algocf}{0}
\setcounter{figure}{0}
\setcounter{table}{0}
\setcounter{equation}{0}

\section{Implementation Details}

We perform control over the seven phases (see Figure~\ref{fig:transition_diagram}) during basketball games through six primitive policies, where defending behaviors are considered as special stances during locomotion and are integrated with a locomotion policy, plus two intermediate policies of gathering for shooting and passing respectively.
All the primitive policies are trained under a multi-objective learning framework for physics-based character control using reinforcement learning~\cite{composite}.
The system architecture for primitive policy learning is shown in Figure~\ref{fig:iccgan}. 
For motion imitation,
we utilize partially observable (hands-only or body-only) reference motions collected from different sources, 
and combine the usage of basketball-playing motions with non-basketball-playing motions for locomotion during dribbling.
In contrast to previous work~\cite{wang2024skillmimic} using detailed full-body motions with ball trajectories for skill learning,
our approach employs a GAN-like architecture~\cite{iccgan}
and enables motion imitation from unstructured sources to achieve more flexible results of motion synthesis without needing pre-collected or generated~\cite{liu2018learning} ball trajectories for reference.
For adaptation to primitive policies during the training for the policy transition,
we use the fine-tuning approach from AdaptNet~\cite{adaptnet}.
This approach adapts a pretrained policy through latent space manipulation,
and allows introducing new context input during adaptation. 

\begin{figure}
    \centering
    \begin{subfigure}[t]{.3\linewidth}
    \includegraphics[width=\linewidth]{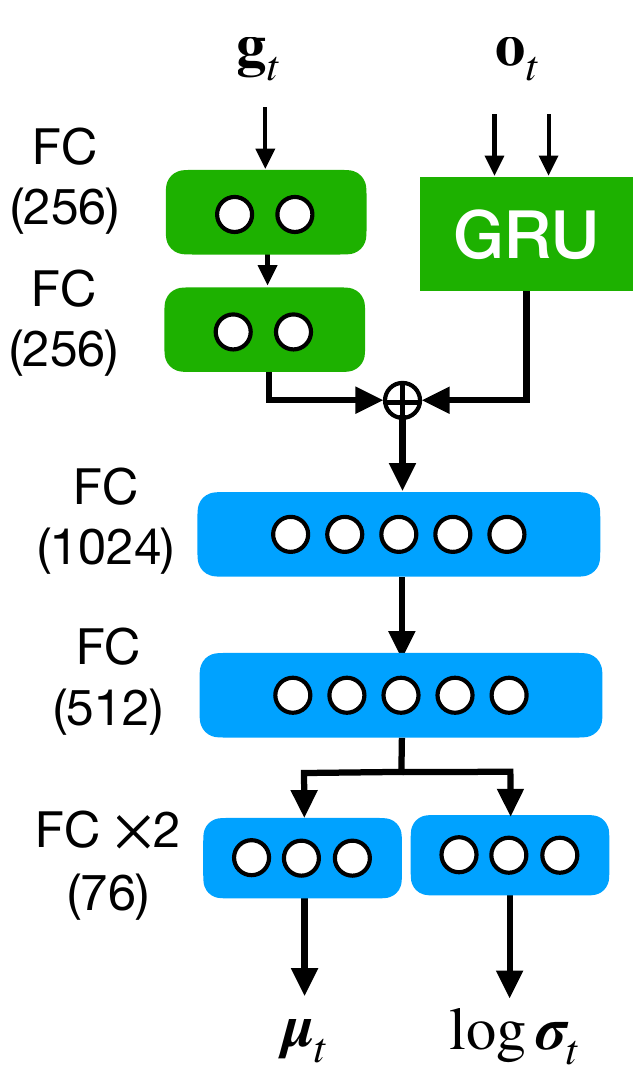}
    \caption{Policy Network}
    \end{subfigure}\quad
    \begin{subfigure}[t]{.3\linewidth}\centering
    \includegraphics[width=\linewidth]{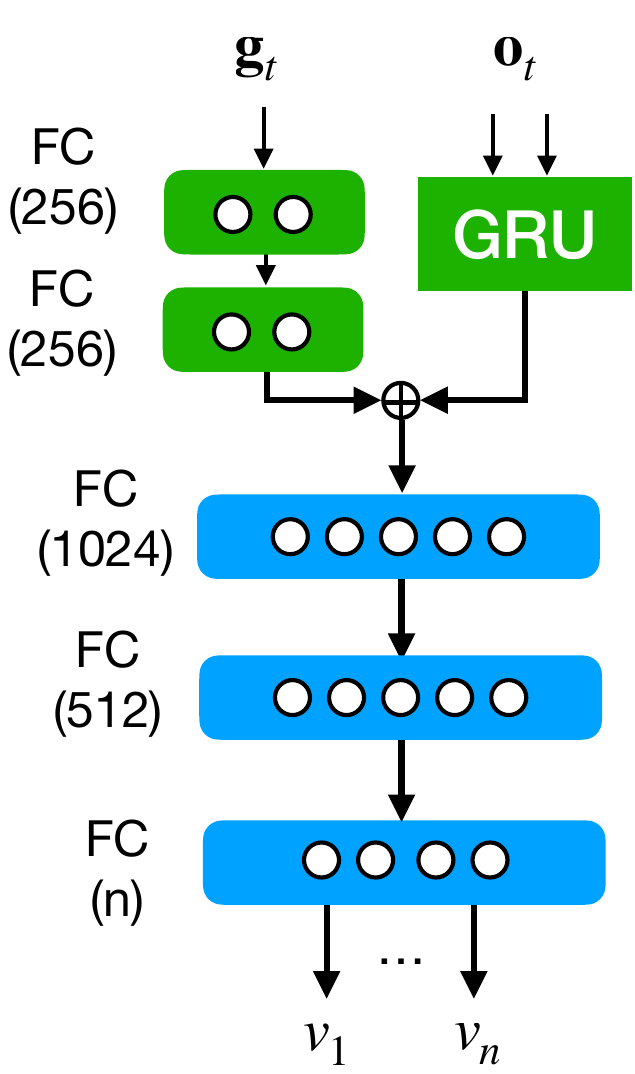}
    \caption{Value Network}
    \end{subfigure}
    \begin{subfigure}[t]{.35\linewidth}\centering
    \includegraphics[width=.81\linewidth]{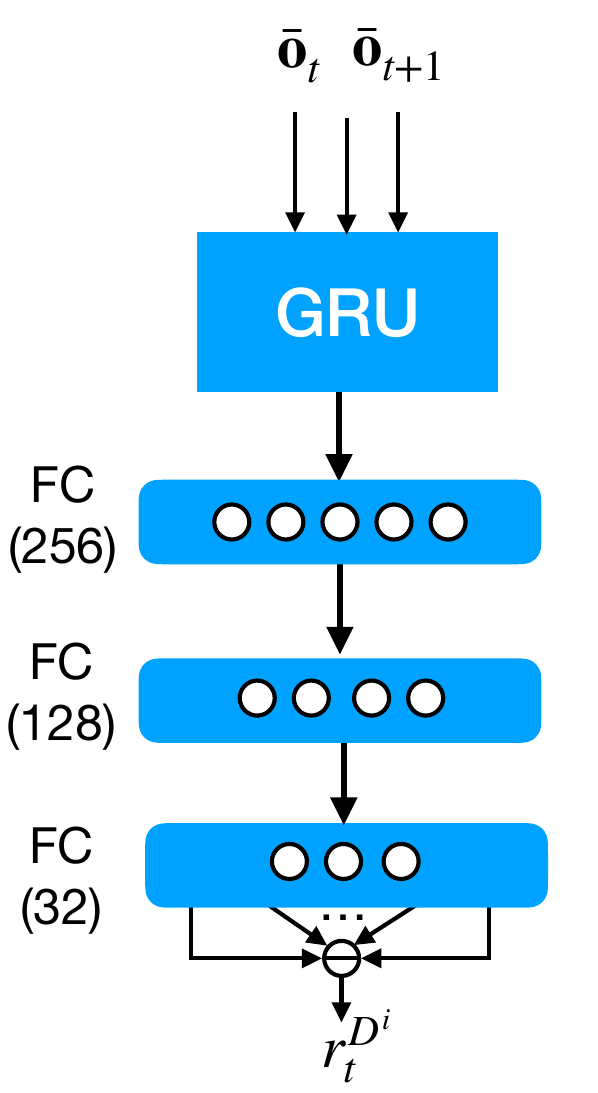}
    \caption{Discriminator Network}
    \end{subfigure}
    \caption{Network structures where the input state $\mathbf{s}_t$ includes $\mathbf{g}_t$ as the goal state input and $\mathbf{o}_t$ as the observation to the character in the last two frames, and $\mathbf{\bar{o}}_t$ and $\mathbf{\bar{o}}_{t+1}$ represents the partial observation to the character in two consecutive frames. We use $\oplus$ denoting the add operator and $\ominus$ denoting the average operator. The state encoder consists of blocks in green, where a GRU encoder is employed for pose state encoding temporally, and a two-layer MLP encoder is employed for goal state encoding. The value network is a multi-head network whose output has a dimension of $n$ depending on the number of objectives in a given task. We have $n=4$ for the dribbling policy, $2$ for passing and shooting policies, and $3$ for the others (cf. Table~\ref{tab:objective_weights}).}
    \label{fig:network}
\end{figure}

We use PPO~\cite{schulman2017proximal} as the backbone reinforcement learning algorithm and take the Adam optimizer~\cite{kingma2014adam} to perform network optimization for policy training.
The hyperparameters used for policy training are listed in Table~\ref{tab:hyper} and the network structures are shown in Figure~\ref{fig:network}.
The optimization function for each primitive policy learning can be written as
\begin{equation}\label{eq:low_level_policy_optimization}
    \max \mathbb{E}_t\left[\sum\nolimits_\kappa w_k \bar{A}_{t,k} \log \pi(\mathbf{a}_t | \mathbf{s}_t) \right]
\end{equation}
where $\bar{A}_{t,k}$ is the standardized advantage that is estimated according to the achieved reward of each objective $k$, and $w_k$ is an associated weight.
The number of objectives and the associated weights differ depending on the given task,
which are summarized in Table~\ref{tab:objective_weights}.

Following previous literature~\cite{iccgan}, the imitation-related reward is obtained through a discriminator ensemble $D$ using hinge loss~\cite{lim2017geometric}:
\begin{equation}\label{eq:dis_rew}
    r_t^{\text{imit}, i}(\mathbf{\bar{o}}_t^i, \mathbf{\bar{o}}_{t+1}^i) = \frac{1}{N} \sum_{n=1}^N \textsc{Clip}\left(D_n^i(\mathbf{\bar{o}}_t^{i}, \mathbf{\bar{o}}_{t+1}^{i}), -1, 1\right),
\end{equation}
where the subscript $i$ indicates different imitation objectives.
Most of our policies have two imitation objectives for hand-only and body imitation (see Table~\ref{tab:objective_weights}).
Therefore,
instead of using a full observation to the character,
we use $\mathbf{\bar{o}}_t^i$ and $\mathbf{\bar{o}}_{t+1}^i$ to represent partially observable states of the character for the corresponding discriminator.
We employ an ensemble of $N$ discriminators ($N=32$ in our implementation) for each imitation objective and the discriminator is trained to minimize the hinge loss with gradient penalty~\cite{gulrajani2017improved}.
We refer to the previous literature~\cite{iccgan} for details.

\begin{table}
\centering
\caption{Hyperparameters. The number of simulated environments would be doubled or tripled during policy transition learning involving multiple primitive policies.}
\begin{tabular}{lc}
    \toprule
    \textbf{Parameter} & \textbf{Value}\\
    \midrule
    policy network learning rate & $5 \times 10^{-6}$\\
    critic network learning rate & $1 \times 10^{-4}$\\
    discriminator learning rate & $1 \times 10^{-5}$\\
    reward discount factor ($\gamma$) & $0.95$ \\
    GAE discount factor ($\lambda$) & $0.95$ \\
    surrogate clip range ($\epsilon$) & $0.2$ \\
    gradient penalty coefficient ($\lambda^{GP}$) & $10$ \\
    number of PPO workers (simulation instances) & $512$ \\
    PPO replay buffer size & $4096$ \\
    PPO batch size & $256$ \\
    PPO optimization epochs & $5$ \\
    discriminator replay buffer size & $8192$ \\
    discriminator batch size & $512$ \\
  \bottomrule
\end{tabular}
\label{tab:hyper}
\end{table}

For the value network,
we employ a multi-head network structure and train it for the multiple objectives of each primitive policy. 
We employ the technique of PopArt~\cite{van2016learning} to perform value normalization on the multiple outputs of a value network.
The normalized state value associated with the task reward is also used for intermediate policy learning (see Section~\ref{sec:middle_level}) or for the preceding policy adaptation in the scene without the intermediate policy (e.g., catching to shooting and catching to passing).

\begin{figure}
    \centering
    \includegraphics[width=\linewidth]{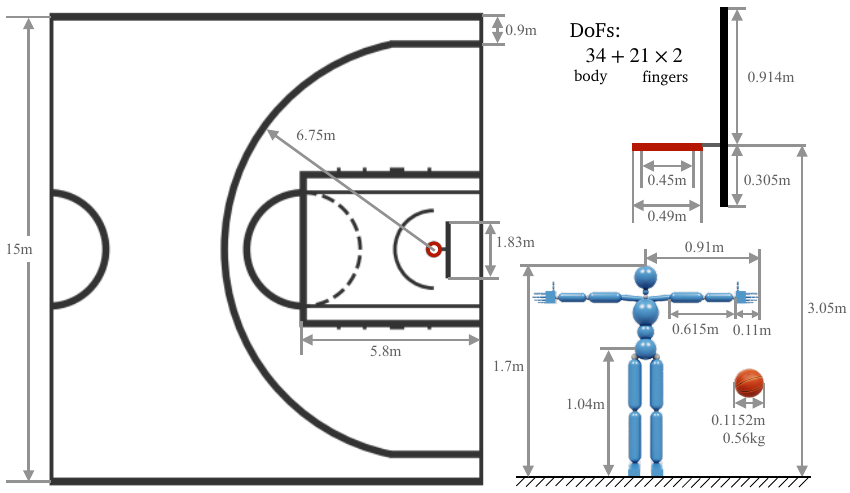}
    \caption{Dimensions of the basketball court and character in our implementation. Our settings follow the standard of Olympic basketball games for women. The character has a normal height and intends to imitate our source motions collected from non-professional basketball players.}
    \label{fig:court}
\end{figure}

Figure~\ref{fig:court} shows the simulated character and the dimensions of the basketball court.
The character has 57 body links and 26 controllable joints with 76 degrees of freedom. 
This results in a state space $\mathbf{s}_t\in\mathbb{R}^{2\times 57 \times 13}$ including the character's position, orientation, and linear and angular velocities of each body link in the last two historical frames.
For ball control,
we only take into account the ball's latest dynamics without historical information and ignore its orientation given its round shape.
This leads to a space of $\mathbb{R}^9$, including the ball's position, and linear and angular velocities, in the goal state vector $\mathbf{g}_t$,
except for the locomotion policy which does not need ball state information.
Depending on the given task,
$\mathbf{g}_t$ could change to include additional information, for example, target velocity for locomotion, hoop positions relative to the character, pivoting foot, and dribbling state of the ball,
which will be elaborated in the following subsections.
All joints in our implementation are controlled through a PD servo.
The policy network outputs the target posture to the PD servo
and leads to an action space $\mathbf{a}_t \in \mathbb{R}^{76}$.
We use IsaacGym~\cite{makoviychuk2021isaac} as our physics simulation engine. All policies run at 30Hz, while the simulation runs at 120Hz.

\begin{table}
    \caption{Objective weights for primitive policy training. A simple weight choice of $0.2$, $0.1$ and $0.7$ (for body imitation, hand imitation, and goal-directed task, respectively) works in most cases. The listed weights are fine-tuned for fast task learning with minimal compromise in motion imitation performance. The rebounding policy use the same weights as the catching policy.}
    \begin{tabular}{rcccccc}
    \toprule
                       & {\small Dribble} & {\small Shoot} & {\small Pass} & {\small Catch} & {\small Gather} & {\small Locomotion}\\
    \midrule
        {\small Body imit.} &  $0.2\ $ & $0.4^\dagger$ & $0.2^\dagger$ & $0.08$ & $0.2$ & $0.25$ \\
        {\small Hand imit.} & $0.1\ $ & - & - & $0.02$ & $0.1$ & $0.15$ \\
        {\small Task} & $0.7^*$
         & $0.6\ $ &$ 0.8\  $ & $0.9$ & $0.7$ & $0.6$ \\
  \bottomrule
  \multicolumn{7}{l}{\footnotesize{$^*$ dynamic weights for navigation and dribbling (see Section~\ref{sec:dribble_policy}).}}\\
  \multicolumn{7}{l}{\footnotesize{$^\dagger$ full-body with hand}}
    \end{tabular}
    \label{tab:objective_weights}
\end{table}

\subsection{Dribbling Policy}\label{sec:dribble_policy}
Besides two imitation objectives, the dribbling policy has two goal-directed rewards for velocity-controlled
navigation and dribbling, respectively.

The navigation reward is defined based on the error between the character's current horizontal velocity $\mathbf{v}$ and the target velocity $\mathbf{v}_\text{target} \in \mathbb{R}^2$: 
\begin{equation}\label{eq:r_nav}
    r_\text{nav} = 
        \exp\left(-\frac{2}{\max\{1, ||\mathbf{v}_\text{target}||^2\}} ||\mathbf{v} - \mathbf{v}_\text{target}||^2\right).
\end{equation}
The target velocity is given to the policy as part of the goal state in the form of velocity direction and magnitude.
During training, the target velocity is generated randomly with a speed sampled in the range between 0 and 4m/s. To help train in-place dribbling motions, there is a chance around 20\% to draw a zero target velocity. 

For dribbling, we perform violation detection to check if there is any invalid contact between the ball and the character's body links other than the hands.
To prevent consecutive dribbling performed in a row before the ball drops on the ground,
we include an additional 0/1 variable $c_\text{dribble}$ with the ball state to indicate if the ball currently should be dribbled or not.
Along with the three goal state variables from the navigation objective, this leads to a goal state $\mathbf{g}_t^\text{dribble} \in \mathbb{R}^{13}$ in total for navigation while dribbling.
The reward function for dribbling is defined as
\begin{equation}\label{eq:r_dribble}
    r_\text{dribble} = 
    0.6r_\text{hand} + 0.4r_\text{sp} + 0.5 I_\text{dribble} (0.2+0.8r_\text{fingers}).
\end{equation}
Specially, $r_\text{dribble} = -1$ if any violation is detected.

\begin{figure}
    \centering
    \includegraphics[width=0.5\linewidth]{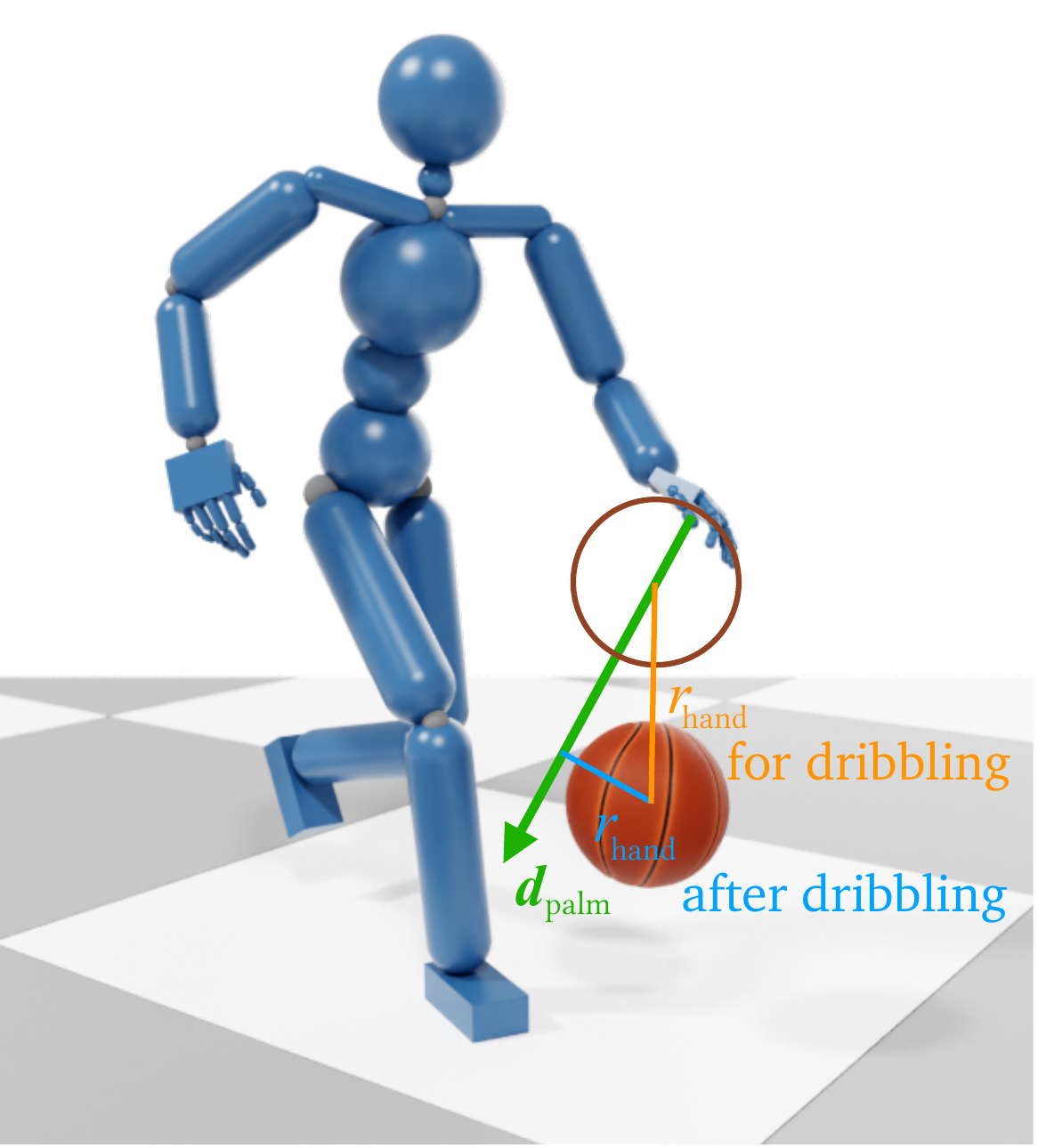}
    \hfill
    \includegraphics[width=0.4\linewidth]{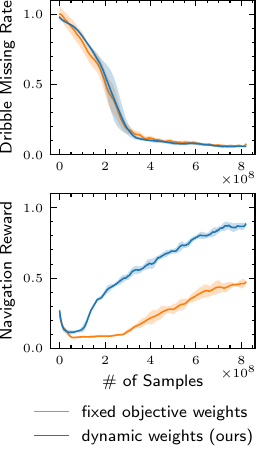}
    \hfill
    \caption{Left: demonstrations of reward computation for ball dribbling before (orange) and after (blue) a dribble happens. Right: comparison of learning performance using dynamic task objective weights and that using fixed objective weights.}
    \label{fig:dribble_rew}
\end{figure}

The term $r_\text{hand}$, as shown in Figure~\ref{fig:dribble_rew}, measures the distance between the character's hand and the ball, encouraging the character to position their hands with palm direction facing toward the ball:
\begin{equation}\label{eq:r_hand}
    r_\text{hand} = \begin{cases}\exp\left(-2||\mathbf{p}_\text{palm}^{\kappa}+R_\text{ball}\mathbf{d}_\text{plam}^{\kappa} - \mathbf{p}_\text{ball}||\right) \\
        \qquad\quad\;\: \text{ if $c_\text{dribble}=1$, i.e., the ball should be dribbled,} \\
    \exp\left(-5||\min_{a>0} \mathbf{p}_\text{palm}^{\kappa}+a\mathbf{d}_\text{plam}^{\kappa} - \mathbf{p}_\text{ball}||^2\right) \text{ otherwise,} \\
    \end{cases}
\end{equation}
where $\mathbf{p}_\text{palm}^{\kappa}$ and $\mathbf{p}_\text{ball}$ are the positions of the palm and ball with $\kappa \in \{\text{left}, \text{right}\}$ to indicate the target hand, $\mathbf{d}_\text{palm}^\kappa$ is the facing direction of the palm, and $R_\text{ball}$ is the radius of the ball.
In the first case, when the ball should be dribbled,
the reward encourages the palm to face the center of the ball while reaching for it,
preventing dribbling with the back of the hand.
In the second case, where the basketball has been dribbled and has not yet rebounded from the ground,
the reward computes the projection of $\mathbf{p}_\text{ball}$ on the palm facing direction $\mathbf{d}_\text{plam}$.
It only encourages the palm to face toward the ball rather than reaching the ball.
The reward in the second case uses a squared error and is more tolerant than the one in the first case using an absolute error. 
It allows the palm to roughly face toward the ball,
while the reward in the first case expects the palm to contact with the ball as closely as possible to generate human-like hand poses to interact with the ball.
The value of $a$ can be found by
\begin{equation}
    a = \max\{0, (\mathbf{p}_\text{ball} - \mathbf{p}_\text{palm}) \cdot \mathbf{d}_\text{plam}\}.
\end{equation}
We decide $\kappa$ by the nearest hand to the ball, i.e.
\begin{equation}
    \kappa = \arg\min_\kappa ||\mathbf{p}_\text{palm}^\kappa - \mathbf{p}_\text{ball}^\kappa||
\end{equation}

The term of $r_\text{sp}$ in Eq.~\ref{eq:r_dribble} measures the vertical speed of the ball and checks if the ball can rebound to at least the height of the pelvis.
\begin{equation}
    r_\text{sp} = \min\{1, |v_\text{ball}/v_\text{target}|\}
\end{equation}
where $v_\text{ball}$ is the current vertical speed of the ball. The target speed $v_\text{target}$ is computed according to the current height of the character's pelvis link $h_\text{pelvis}$.
It takes into account the restitution coefficient $e$ between the ball and the ground to compute the rebound height if the ball is moving toward the ground vertically (i.e., $v_\text{ball}<0$),
or consider only the impact from gravity if the ball is moving upward:
\begin{equation}
    v_\text{target} = \begin{cases}
        \sqrt{-2g(h_\text{pelvis} - h_\text{ball})} & \text{if $v_\text{ball} > 0$,}\\
        \sqrt{-2g(h_\text{pelvis}/e^2 - h_\text{ball})} & \text{otherwise,}
    \end{cases}
\end{equation}
where $g=-9.81m/s^2$ is the gravitational acceleration and $e=0.875$ is the restitution coefficient between the foot and ground.
When $h_\text{pelvis} < h_\text{ball}$ or $h_\text{pelvis}/e^2 < h_\text{ball}$, we have $r_\text{sp} = 1$.
The restitution coefficient is decided by measuring the bounce height of the ball, according to the basketball rules, in the physics simulator. It takes into account the simulation error and is slightly different from the physical quantity in the real world.
In our experiments,
we find that the introduction of $r_\text{sp}$ is crucial for the policy to master continuous dribbling behaviors.

\begin{figure*}
    \centering
    \includegraphics[width=\linewidth]{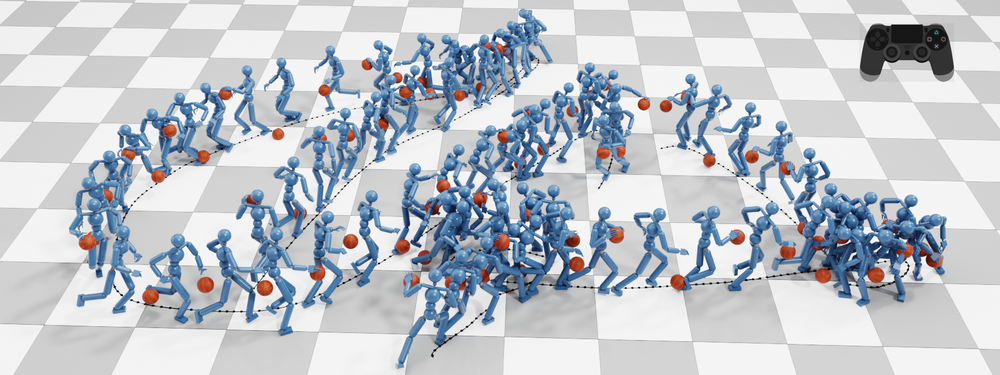}
    
    \includegraphics[width=.075\linewidth]{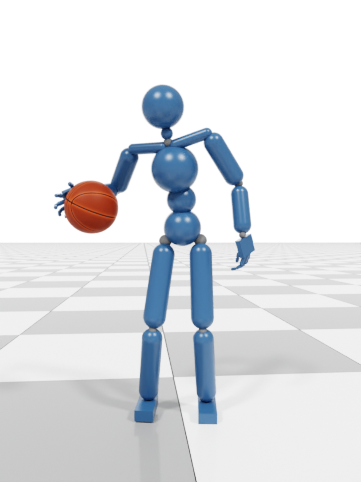}
    \includegraphics[width=.075\linewidth]{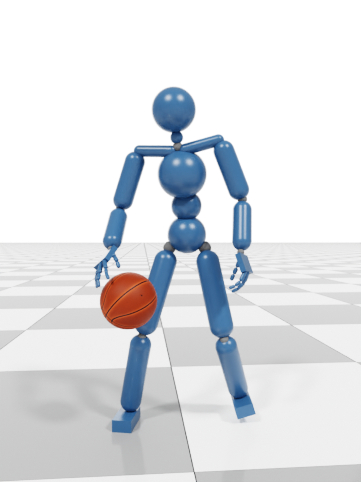}
    \includegraphics[width=.075\linewidth]{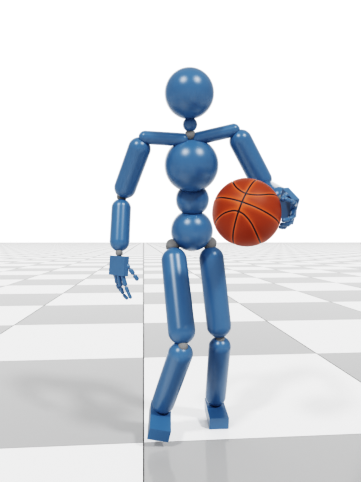}
    \hfill
    \includegraphics[width=.075\linewidth]{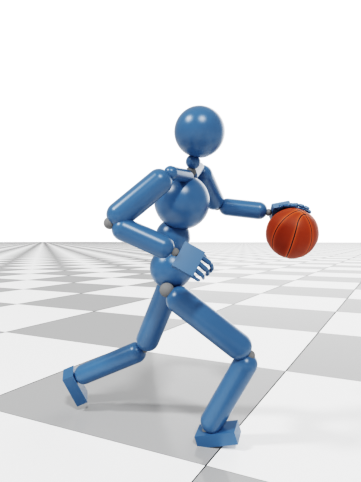}
    \includegraphics[width=.075\linewidth]{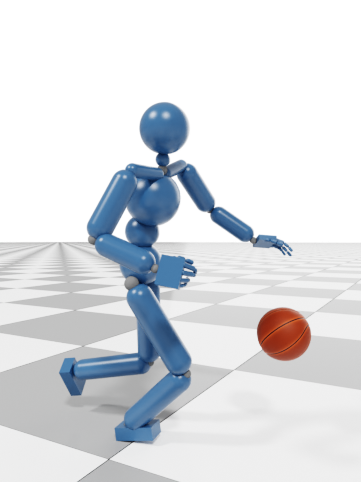}
    \includegraphics[width=.075\linewidth]{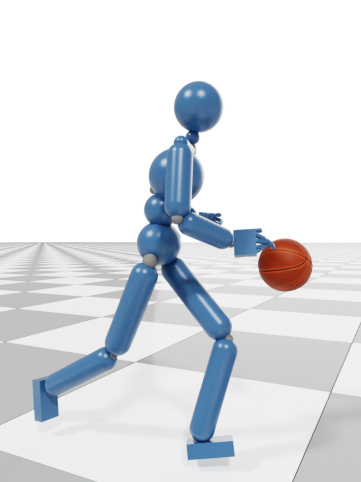}
    \hfill
    \includegraphics[width=.075\linewidth]{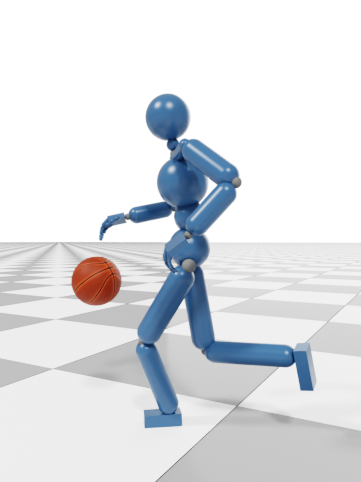}   \includegraphics[width=.075\linewidth]{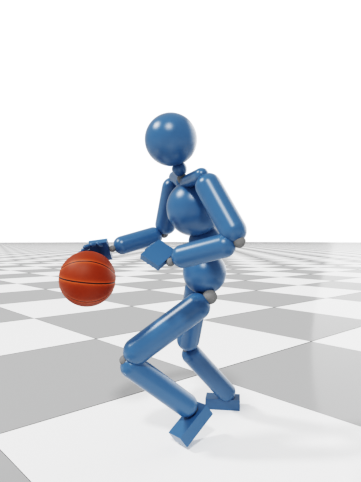}
    \includegraphics[width=.075\linewidth]{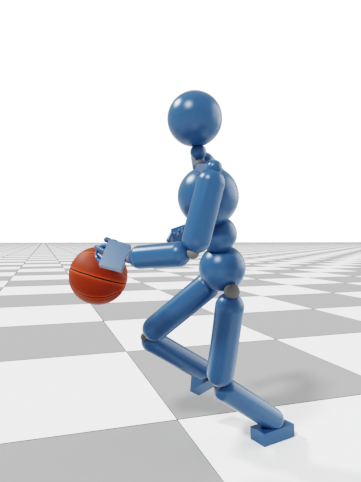}
    \hfill
    \includegraphics[width=.075\linewidth]{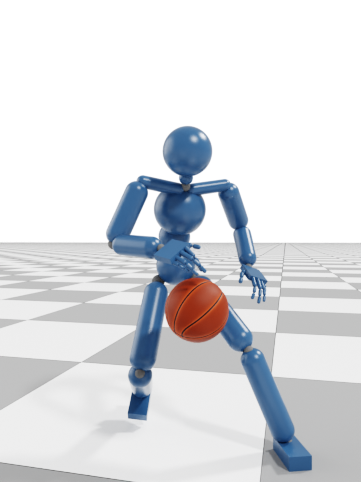}
    \includegraphics[width=.075\linewidth]{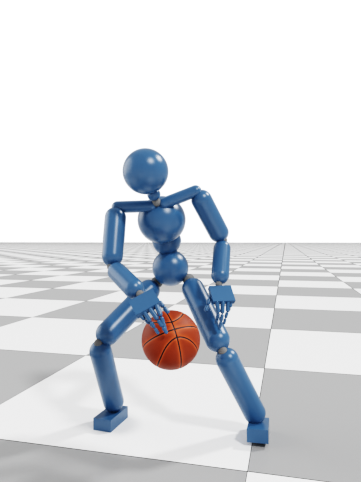}
    \includegraphics[width=.075\linewidth]{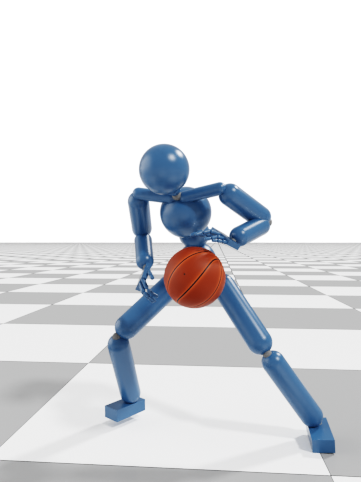}
    \caption{Demonstrations of our dribbling policy trained using partially observable motion data. The dribbling policy is trained with unstructured motion data collected from multiple sources without ball trajectories. The character controlled by the policy can interact with the ball validly using hands and support interactive control by giving arbitrary target velocities with a valid speed in the range between 0m/s (inclusive) and 5m/s. Bottom: close-up pictures captured during the character walking, running and in-place dribbling.}
    \label{fig:dribble}
\end{figure*}

\begin{algorithm}[t]
\SetKwInOut{Input}{Input}
\SetKwInOut{Output}{Output}
\nl\uIf{$v_\text{ball}^t - v_\text{ball}^{t-1} < \Delta V - 0.01$ and $v_\text{ball}^{t} < 0$}{
\nl    $I_\text{dribble} = 1$
\tcp*{A dribble happens.}
\nl    $c_\text{dribble} = 0$ \tcp*{Another dribble should not start.}
}
\nl\uElse{
\nl    $I_\text{dribble} = 0$
}
\nl\uIf{$v_\text{ball}^{t-1} < 0$ and $v_\text{ball}^t >0$}{
\tcp{Ball bounces up from the ground.}
\tcp{Count the dribble missing rate.}
\nl\uIf{$c_\text{dribble} = 1$}{
\nl $n_\text{dribble}^- = n_\text{dribble}^- + 1$ \tcp*{Missing one dribble.}
}
\nl\uElse{
\nl $n_\text{dribble}^+ = n_\text{dribble}^+ + 1$
}
\nl    $c_\text{dribble} = 1$ \tcp*{Another dribble is allowed.}
}
\caption{Ball Dribbling State Detection}
\label{alg:dribbling}
\end{algorithm}

The term of $I_\text{dribble} r_\text{fingers}$
in Eq.~\ref{eq:r_dribble} measures the finger pose when dribbling happens, where
$I_\text{dribble}$ is a 0/1 variable to indicate if a dribble happens. It is equal to 1 if the ball is pushed downward by a hand. The definition of $r_\text{fingers}$ can be written as
\begin{equation}\label{eq:r_fingers}
    r_\text{fingers} = 
    \exp\left(-10 \sum_{f\in\mathcal{F}^\kappa} |\mathbf{p}_\text{f}^\kappa -\mathbf{p}_\text{ball}|-R_\text{ball} \right)
\end{equation}
where $\mathcal{F}^\kappa$ is the set all fingertips plus the palm of the hand $\kappa$,
and $\mathbf{p}_\text{f}^\kappa$ is the global position of a fingertip or the palm center in 3D space.
This reward encourages the hand to dribble the ball with all fingertips in contact with the ball.

Instead of relying on accurate contact information,
we use a simple heuristic rule to decide the ball's dribbling states, as shown in Algorithm~\ref{alg:dribbling},
where $\Delta V = g/\text{FPS}$ is the velocity change of free-falling objects in the physics simulator, given that IsaacGym uses Euler's method for integration.

Since we do not have references of ball trajectories, the ball is initialized in front of the character randomly with a distance from 0.5 to 0.8m.
If the character falls down, a termination reward of $-25$ will be given, and the simulation episode terminates.
For invalid dribbling behaviors or violations, however,
we adopt a soft termination strategy to
re-initialize the position of the ball in front of the character and assign a penalty reward of $-1$ to the policy,
while the simulation episode does not terminate.
By such,
we avoid the policy termination occurring too often at the beginning of training, which could prevent the policy from even learning to walk.

Additionally,
we take a dynamic weighting strategy for the dribbling policy training.
Since dribbling is much more difficult than locomotion and there is a potential conflict between the two tasks,
to balance the two task objectives,
a simple solution is to assign a larger weight in Eq.~\ref{eq:low_level_policy_optimization} to the dribbling task, given its higher priority.
However, this approach can negatively impact the learning of locomotion.
To address this issue,
we set a lower weight to the locomotion objective initially to let the policy learn more about dribbling at the beginning of training, but increase it progressively as the success rate of dribbling ($p_\text{dribble}$) increases, namely, the dribble missing rate ($1 -p_\text{dribble}$) decreases.
In our implementation,
$p_\text{dribble}$ is counted by moving average based on the policy's performance in training rollout (8 simulation steps with 512 parallel environments) with a fading coefficient of $0.9$ on the past performance, i.e.
\begin{equation}
    p_\text{dribble} \leftarrow 0.9 p_\text{dribble} + 0.1\frac{n_\text{dribble}^+}{n_\text{dribble}^- + n_\text{dribble}^+}
\end{equation}
where $n_\text{dribble}^-$ and $n_\text{dribble}^+$ are the number of counted successful and missed dribbles (cf. Algorithm~\ref{alg:dribbling}).
The weight for the locomotion objective is set dynamically and non-decreasing during training based on the dribble-missing rate through
\begin{equation}
    w_\text{nav} \leftarrow 0.2+0.5\max\left\{\exp\left(-10(1 -p_\text{dribble})\right), w_\text{nav}\right\}
\end{equation}
with an initial value of $w_\text{nav}=0.2$,
and the weight for the dribble objective is set by
\begin{equation}
    w_\text{dribble} = 0.7 - w_\text{nav}.
\end{equation}
The two goal-directed tasks have a total weight of 0.7, and the total weight of the two motion imitation objectives is 0.3.
In Figure~\ref{fig:dribble_rew},
we compare the learning performance using our dynamic weight strategy and that using fixed weights (0.2 for navigation and 0.5 for dribbling, i.e., the initial weights for the dynamic weight strategy). 
From the figure, we can see that
as the dribble missing rate decreases,
there is no performance loss when we reduce the weight of dribbling (blue).
However, the increase in the associated locomotion weight makes the learning of navigation more effective.

\subsection{Shooting Policy}
\begin{figure}
    \vspace{-1.5em}
    \centering
    \includegraphics[width=.32\linewidth]{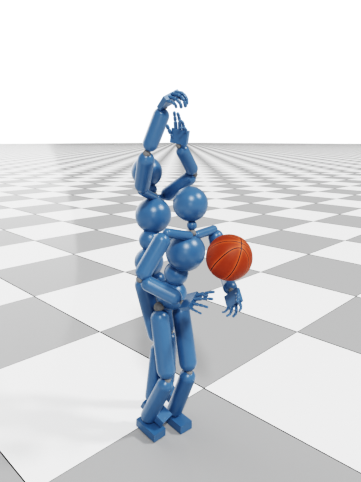}
    \includegraphics[width=.32\linewidth]{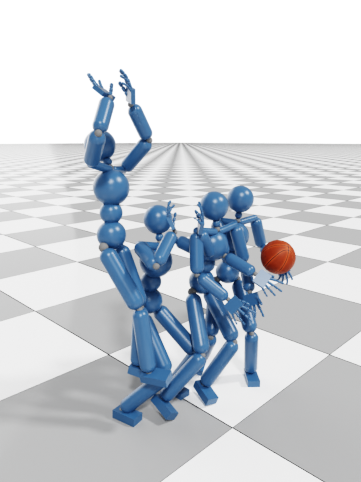}
    \includegraphics[width=.32\linewidth]{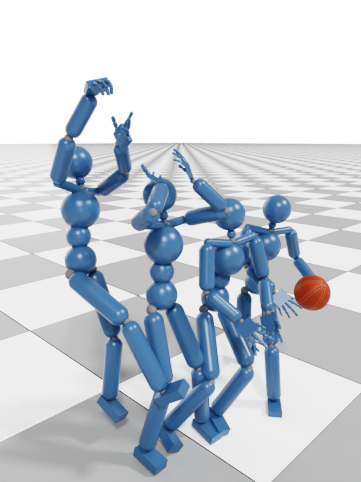}
    
    \includegraphics[width=.32\linewidth]{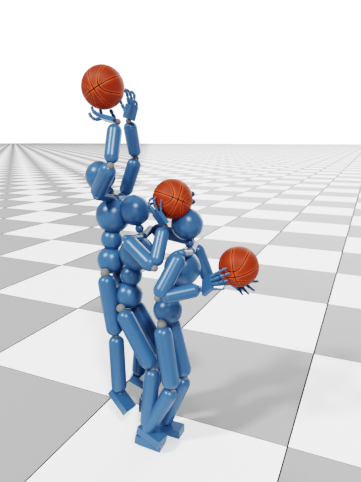}
    \includegraphics[width=.32\linewidth]{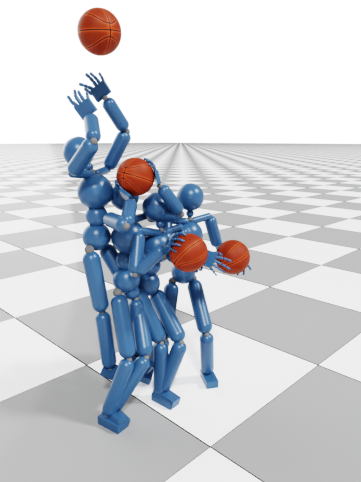}
    \includegraphics[width=.32\linewidth]{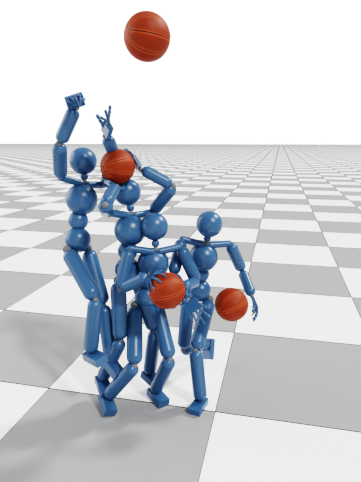}
    \caption{Top: reference motions of jump shooting without ball trajectories but a manually decided initial position of the ball. Bottom: close look of the learned shooting motions. }
    \label{fig:shoot_ref_motions}
\end{figure}

For the shooting policy, we have
a full-body (including hands) imitation objective and a goal-directed reward that measures the shot accuracy and ball-holding performance before the ball is shot out.
Besides the ball state, the shooting policy also takes the horizontal position of the hoop relative to the character as the goal state.
This leads to a goal state vector $\mathbf{g}_t^\text{shoot} \in \mathbb{R}^{11}$.

Due to missing ball trajectory reference, 
we manually set the ball's initial position near the character’s hand in each of the three reference motions that we have for policy learning,
as shown in Figure~\ref{fig:shoot_ref_motions}.
This initialization scheme makes the ball easily catchable for the character through motion imitation and avoids unexpected penetrations between the ball and character caused by random initialization.
Nevertheless, such an initialization scheme lets the trained policy see only a very limited set of the ball states and thus makes it sensitive to the initial pose of the ball during execution.
Each training episode will end after 60 frames (2s) and terminate if the ball has not been released with two hands holding the ball after 40 frames.

The reward for shooting encourages the character to first hold and then lift the ball using two hands before shooting the ball out, like a human player would do for a jumping shoot:
\begin{equation}\label{eq:r_shoot}
    r_\text{shoot} = \begin{cases}
        0.5 r_\text{hands} + I_\text{up} r_\text{hold} \quad\;\: \text{ before the ball is released,} \\
        r_\text{release} + \exp(-0.25 || \mathbf{\hat{p}}_\text{ball} - \mathbf{p}_\text{hoop} ||) \quad\text{ otherwise.}
    \end{cases}
\end{equation}

Similar to the first case in Eq.~\ref{eq:r_hand},
$r_\text{hands}$ encourages the hand to approach the ball with the palm facing toward the ball.
However, while dribbling uses only one hand,
$r_\text{hands}$ here takes into account the two hands at the same time:
\begin{equation}\label{eq:r_hands}
    r_\text{hands} = \exp\left(\frac{-5 }{|\kappa|}\sum_\kappa |\mathbf{p}_\text{palm}^\kappa+R_\text{ball}\mathbf{d}_\text{palm}^\kappa - \mathbf{p}_\text{ball}| \right)
\end{equation}
where $\kappa \in \{\text{left}, \text{right}\}$ and $|\kappa|=2$.
$I_\text{up} = 1$ if the ball is lifted up compared to previous frames or $0$ otherwise.
$r_\text{hold}$ is similar to Eq.~\ref{eq:r_fingers}, but measures the performance of two hands holding the ball simultaneously.
The definition of $r_\text{hold}$ can be written as
\begin{equation}
    r_\text{hold} = \exp\left(-20\sum_\kappa\sum_{f\in\mathcal{F}^\kappa} |\mathbf{p}_\text{f}^{\kappa} - \mathbf{p}_\text{ball} - R_\text{ball}|\right).
\end{equation}
By multiplying $r_\text{hold}$ by $I_\text{up}$,
we expect that the character could lift the ball with fingers and palms in close contact with the ball before shooting it out.

\begin{figure}
    \centering
    \includegraphics[width=\linewidth]{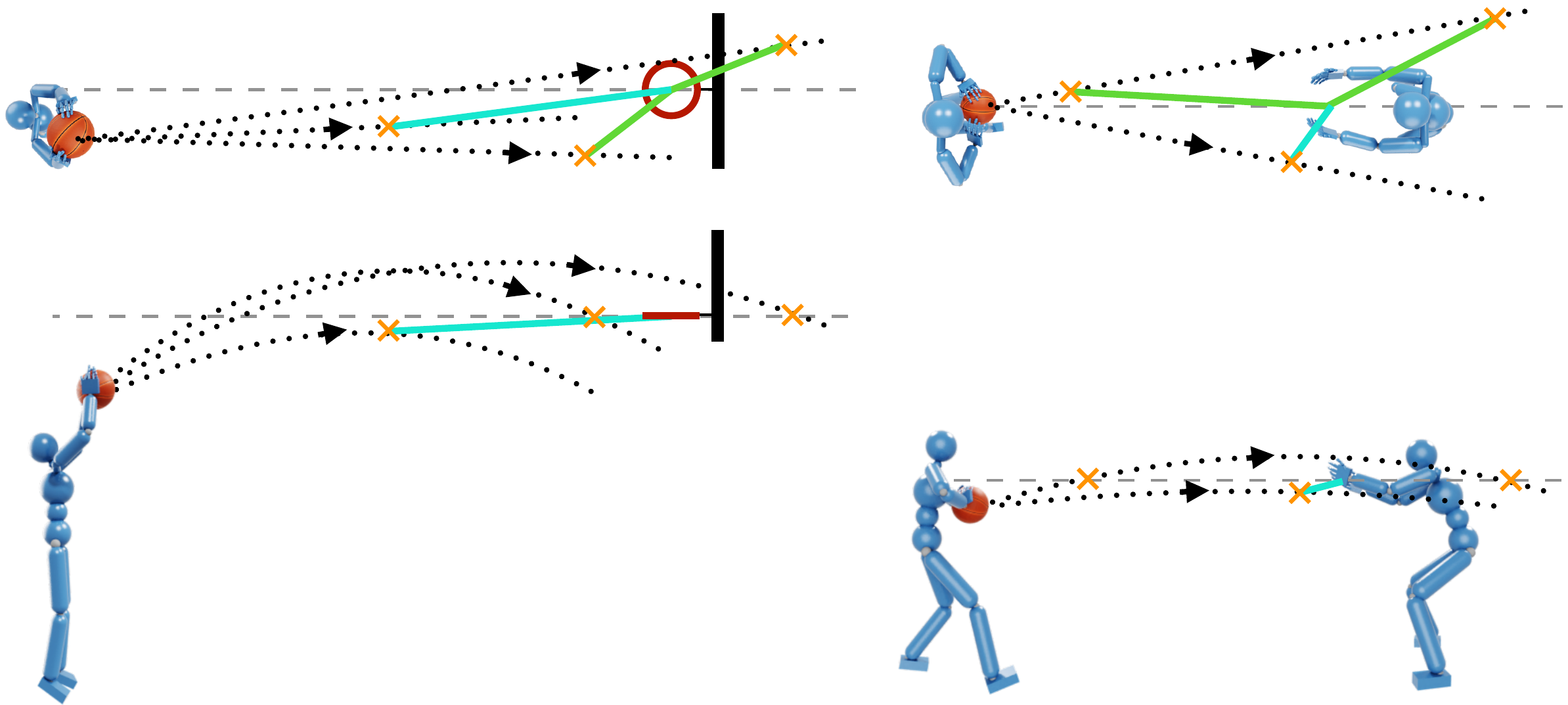}
    \caption{The chosen points (yellow cross markers) on the ball projectile used for reward computation for shooting (left) and passing (right) tasks. We compute the 2D distance from the falling points on the horizontal plane of the target (blue lines) or the 3D distance (cyan) from the top of the projectile if the projectile is not high enough to reach the target height. For the shooting task, we consider the falling point only. For the passing task, we consider the two possible points where the projectile intersects with the plane at the target height.}
    \label{fig:pass_shoot_rew}
\end{figure}

In the second case of Eq.~\ref{eq:r_shoot},
$r_\text{release}$ is defined as 
\begin{equation}
    r_\text{release} = h_\text{release}/h_\text{hoop}
\end{equation}
where $h_\text{release}$ is the height of the ball when it is held by the hand just before being released.
This reward term encourages the character to shoot the ball from a higher position rather than directly throwing the ball out.
In the last term,
$\mathbf{\hat{p}}_\text{ball}$ is the estimated minimal distance between the ball flying trajectory and the hoop on the horizontal plane at the hoop height.
Figure~\ref{fig:pass_shoot_rew} shows how the points are chosen
given the ball's projectile decided by its current linear velocity.
We estimate $\mathbf{\hat{p}}_\text{ball}$ under the assumption that the ball is only affected by gravity, and then correct the estimation based on the actual simulated result of the ball state.

Though the initial posture is fixed,
the character will be randomly put in a ring area 
around the hoop with an inner radius of 2.5m and an outer radius of 7.5m, as shown in Figure~\ref{fig:shoot}.
The initial facing direction will also be adjusted with some randomness to ensure that the character can face the hoop roughly (with a maximal error of $20^\circ$).
As discussed in the experiment section,
the shooting policy is sensitive to the initial posture and ball state,
but can perform shots with high accuracy in the effective area.

We do not perform violation detection when training the vanilla, primitive policy of shooting.
The major violation that could happen during the phases from ball catching to jump shooting is traveling, 
which barely occurs with effective full-body imitation of the reference motions starting with a jump-ready pose.
However, during adaptation learning for policy transition from dribbling or catching,
traveling could happen very frequently,
given the character dynamic states.
In that case,
we introduce an additional variable in the goal state vector to indicate the pivoting foot for the policy to avoid traveling.
The details of traveling detection will be elaborated in Section~\ref{sec:traveling}.

\subsection{Passing Policy}
The passing task is very similar to shooting
with the major difference in (1) the goal state and (2) the falling points chosen for reward computation.
The goal state for passing includes the ball state and the target position of passing in the 3D space relative to the character's root link.
This leads to a goal state $\mathbf{g}_t^\text{pass}\in\mathbb{R}^{12}$.
Similar to the training of the shooting policy, an additional indicator of the pivoting foot will be introduced into the goal state during adaptation learning for policy transition, but not in the pre-training phase of the primitive policy.

While during reward computation, the shooting task only takes the valid falling point where the ball drops from a higher position above the hoop, the passing task allows the ball to reach the target position from any direction (see Figure~\ref{fig:pass_shoot_rew}).
Ideally, we prefer to find the closest position on the ball's projectile to the target position for reward computation.
However, this will lead to an optimization problem involving a cubic equation, which is not easy to solve.
Therefore, instead of finding the closest point in the 3D space,
we adopt a similar strategy of reward computation for the shooting policy, but consider the closest one from the two possible intersection points of the ball's projectile and the horizontal plane of the target position for reward computation.

During training, the position of the passing target is chosen randomly in a half ring area in front of the character with an inner radius of 2.5m and an outer radius of 7.5m at a random height between 0.8 and 1.1m.
During deployment for interactive control,
the chest position of the target character is chosen as the target position of passing.

\subsection{Catching Policy}
The catching policy has the same goal state with the passing policy,
but the indicator variable for the pivoting foot is always introduced to prevent traveling after ball catching.
The goal state, therefore, is $\mathbf{g}_t^\text{catch} \in \mathbb{R}^{13}$.
We refer to Section~\ref{sec:traveling} for details of traveling and pivoting foot detection.

The catching reward is similar to the ball-holding reward used for the shooting policy:
\begin{equation}\label{eq:r_catch}
    r_\text{catch} = 0.5 r_\text{hands}^\prime + r_\text{hold} - I_\text{traveling}
\end{equation}
where $I_\text{traveling}$ is a 0/1 indicator for foot traveling.
The term $r_\text{hands}^\prime$ is an extend to $r_\text{hands}$ in Eq.~\ref{eq:r_hands}:
\begin{equation}
    r_\text{hands}^\prime = 0.3\exp(-e_\text{hands}) + 0.7 r_\text{hands}
\end{equation}
where $e_\text{hands} = \frac{1}{|\kappa|}\sum_\kappa |\mathbf{p}_\text{palm}^\kappa+R_\text{ball}\mathbf{d}_\text{palm}^\kappa - \mathbf{p}_\text{ball}|$.
As the distance from the hands/fingers to the ball may vary a lot during the catching process,
we take the trick from previous literature~\cite{guitar}, and use 
the sum of two exponential functions for reward computation.
This trick prevents it from falling into the saturation range of an exponential function when the error is large, while keeping the reward function sensitive when facing small errors.

\begin{figure}
    \centering
    \includegraphics[width=\linewidth]{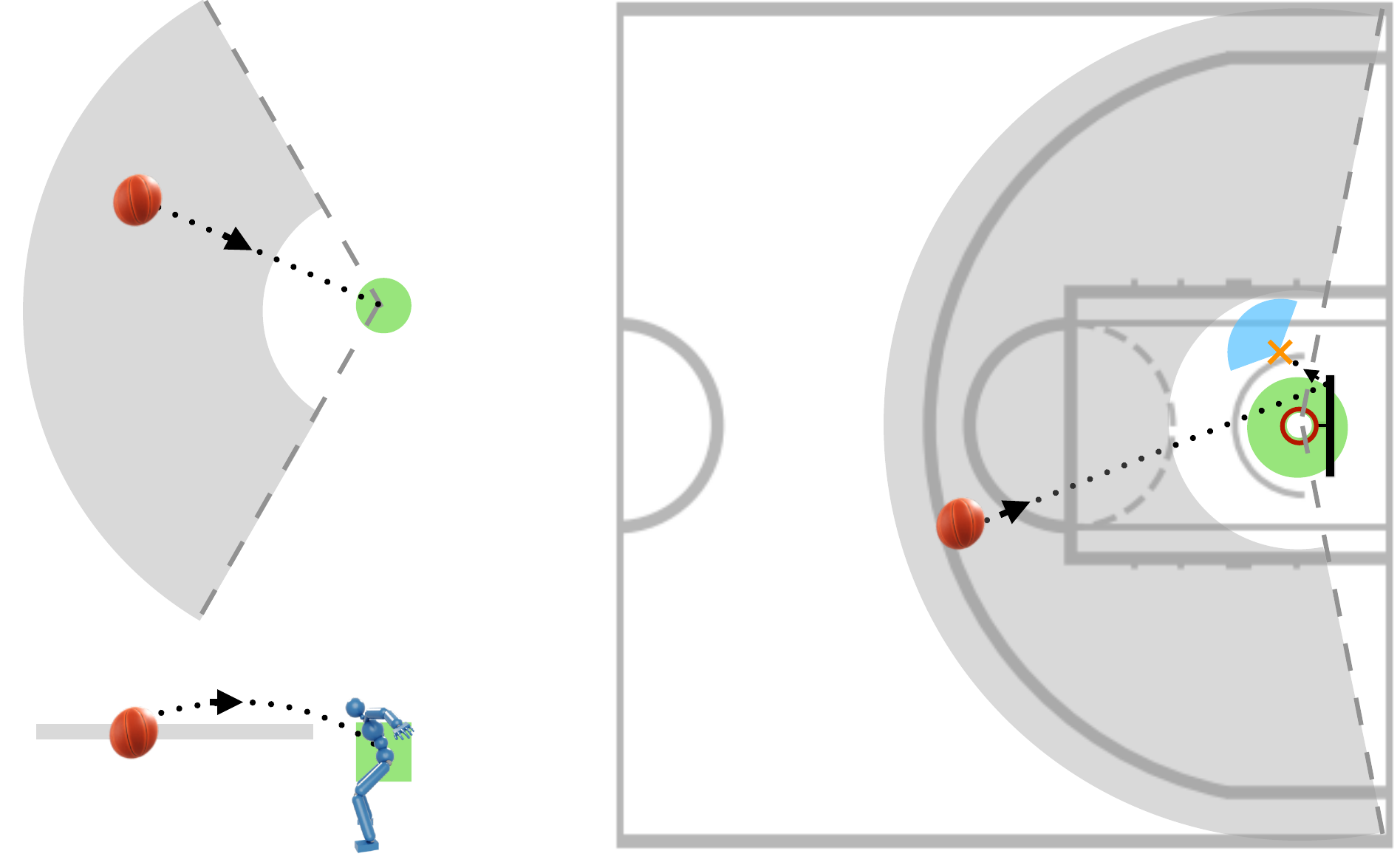}
    \caption{Ball initialization for the training of the catching (left) and rebounding (right) policies. The gray areas indicate the area where the ball will be launched randomly. A falling point in the green area will be chosen randomly to decide the initial velocity of the ball. The falling point is chosen around the catching player while training the catching policy. During rebounding policy training, the character will be randomly placed at the front direction (blue area) of the ball's actual falling point (yellow cross marker) after colliding, if there is, with the hoop and board.}
    \label{fig:catch_rebound_rew}
\end{figure}

To improve the training efficiency,
the ball during training is initialized to fly towards the character roughly, such that the character can try to catch the ball effectively.
Figure~\ref{fig:catch_rebound_rew} shows the area in gray from which the initial position of the ball is randomly drawn.
It is a ring arc area in front of the character with an inner radius of 2.5m and an outer radius of 7.5m.
The angle of the arc is $120^\circ$.
The initial height is randomly chosen from 0.8 to 1.1m.
The target position is randomly sampled from a circular area about the character with a radius of 0.5m  (green in the figure).
The target height is -0.1m to 0.5m around the root height of the character.
We consider the ball's traveling time as a random number between 0.3s (10 frames) and 0.83s (50 frames).
The initial linear velocity of the ball is then decided by the sampled initial position,  target position, and traveling time.

\subsection{Rebounding Policy}\label{sec:rebound}
The rebounding policy is similar to catching.
They are trained using exactly the same reference motions and goal-directed reward in Eq.~\ref{eq:r_catch}.
However, while the catching policy faces the scenario where the ball flies in the air barrier-freely,
the rebounding policy faces the scenario where the ball would collide with the hoop and/or board and change its trajectory.
To enable the policy to detect the potential impacts from the collision,
we use the goal state from the shooting policy, which includes the hoop's position besides the ball's state.
Similarly to the training of the catching policy,
we also introduce the pivoting foot indicator in the goal state to prevent foot traveling.

Figure~\ref{fig:catch_rebound_rew} shows how the ball is initialized for rebounding during training.
To emulate the pre-rebounding states of the ball,
we run additional simulation environments in parallel to collect ball trajectories with falling points near the hoop.
We procedurally shoot the ball from the common shooting area (the gray area in the figure) at a random height between 1.5m and 3m to the hoop range, as shown by the green area in the figure, excluding the center of the hoop.
After collecting the falling trajectories of the ball, which may collide with the hoop and/or board or just free-fall,
we initialize the character in front of the falling point (the blue area).
The facing direction of the character is roughly toward the ball's horizontal direction of velocity at a randomly sampled angle within $120^\circ$ and a distance within 1m.
We also constrain the ball's maximal launching speed to 3m/s horizontally and angle to $45^\circ$.
Since the whole flying trajectory of the ball may be quite long, taking at most 2 seconds,
we cut the trajectory, and extract the ball pose in the last 20th frame (2/3 seconds) before the ball drops at the hoop height as the ball pose initialized together with the character for policy training.

\begin{figure}
    \centering
    \includegraphics[width=\linewidth]{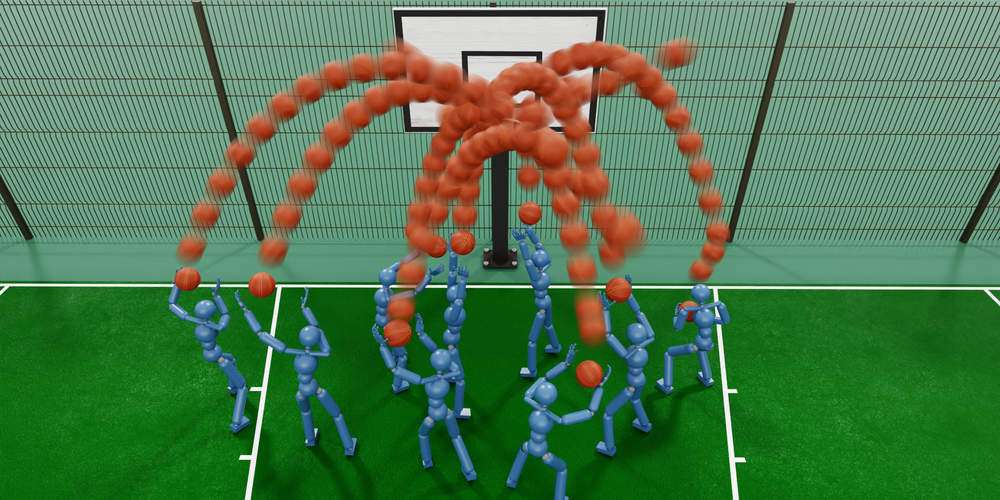}
    \caption{Character controlled by our rebounding policy can perform rebounding in arbitrary positions around the hoop near the falling point.}
    \label{fig:rebound}
\end{figure}

Instead of mimicking certain pre-collected motions to catch the ball with a fixed set of dynamic states,
our trained policy can control the character to perform rebounding in any valid position around the ball's falling point given an arbitrary initial state.
Figure~\ref{fig:rebound} exhibits some results of our rebounding policy.
The character under control can, in advance, predict the falling point of the ball after colliding with the hoop and/or board,
and adjust its pose for better rebounding.
The introduction of foot traveling detection prevents the character from randomly moving after catching the ball.
We refer to our supplementary video for the animated results.

\subsection{Gathering Policy}

The gathering policy is not a completely independent policy.
It is introduced as the intermediate policy for transition purposes between dribbling and shooting, as described in Section~\ref{sec:middle_level}.
Besides the shooting off the dribbling,
we also adopt the same setup to train a gathering policy for passing off the dribbling in our implementation.
Except for the common part of the ball state,
the goal state differs depending on the succeeding policy.
In the shooting-off-the-dribbling task,
the goal state has the hoop position like the shooting policy but with one more indicator for the pivoting foot to prevent traveling.
In the passing-off-the-dribbling task,
the goal state is the pivoting foot indicator plus the position of the passing target used for the passing policy.

The reward function defined for the gathering policy 
is similar to the reward for the shooting policy, but focuses only on the ball-catching behaviors without the ball-releasing reward.
Meanwhile, for policy transition purposes (cf. Eq.~\ref{eq:rgather}),
we utilize the state value function from the adapted shooting policy (i.e. $\pi_\text{shoot}^+$) to guide the character reaching a state preferred by the shooting policy.
A similar approach is used for the passing-off-the-dribbling task while adapting the passing policy.
We refer to Eq.~\ref{eq:rgather} for the full reward definition,
where
\begin{equation}
    r_\text{pose} = 0.3 r_\text{hands}^\prime + 0.2r_\text{orient} + r_\text{hold} - I_\text{traveling}.
\end{equation}
This reward is similar to the reward function for the catching policy~(Eq.~\ref{eq:r_catch}) but has one more term $r_\text{orient}$ measuring the facing direction error of the character.
$r_\text{orient}$ is computed by
\begin{equation}
    r_\text{orient} = \exp\left(-4\left(\frac{ |\arccos \theta|}{\pi}\right)^3 \right)
\end{equation}
where $\theta$ is the angle between the character's pelvis (root) heading direction and the horizontal direction from the character to the hoop in the shooting-off-the-dribbling task and to the target position in the passing-off-the-dribbling task.

During the gathering policy training,
a dribbling policy runs parallelly in another batch of simulation environments to produce random initial states of the character and ball for the gathering policy.
While taking the produced local dynamics of the character and ball,
the character controlled by the gathering policy will be teleported to a random position near the hoop in the same range as the shooting policy training to perform ball gathering for shooting in the shooting-off-the-dribbling task.
Based on the estimated state value, we pass states of the gathering character and ball to the succeeding policy for further adaptation.
We will terminate the episode without penalty if the ball is out of control (flying away or rolling/bounding on the ground) within 40 frames~(1.3s).
A termination penalty (a reward of $-25$) will be given only when the character falls down.

\subsection{Defending and Locomotion Policy}\label{sec:locomotion}
\begin{figure}
    \centering
    \includegraphics[width=\linewidth]{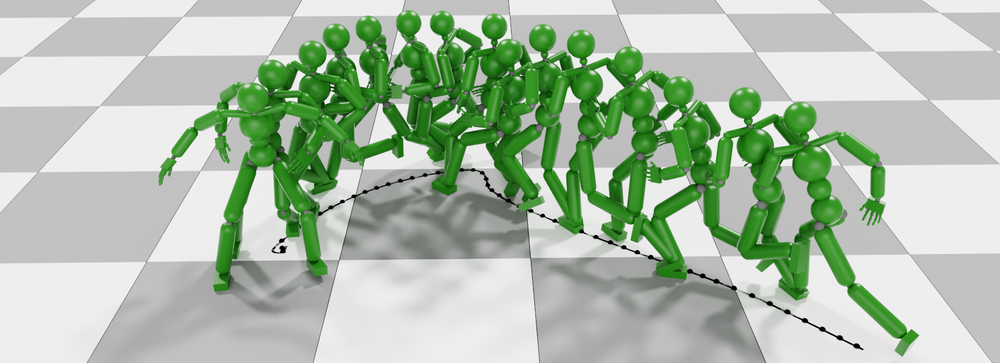}

    \includegraphics[width=0.32\linewidth]{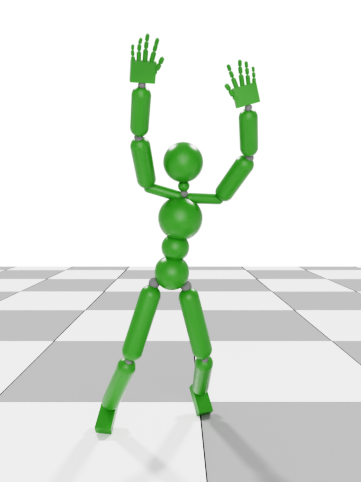}
    \includegraphics[width=0.32\linewidth]{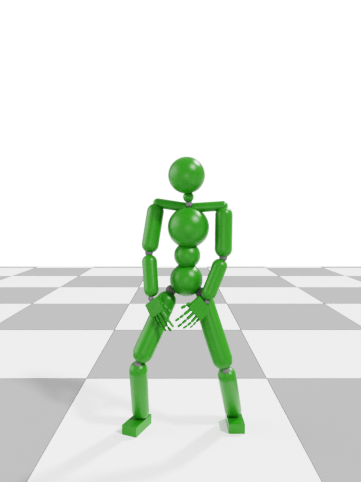}
    \includegraphics[width=0.32\linewidth]{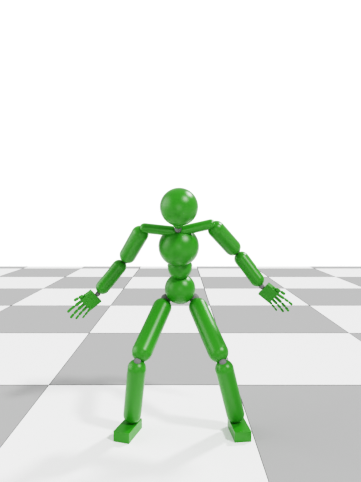}
    \caption{Top: controllable locomotion with defensive stances. Bottom: Close look at defensive behaviors  
    corresponding to different arm stretching poses. From left to right: block, screen, and defensive stance.}
    \label{fig:defensive_locomotion}
\end{figure}

We combine the locomotion and defending behavior into one policy.
The character is expected to perform locomotion when a horizontal target velocity ($\mathbf{v}_\text{target}\in\mathbb{R}^2$) is given,
or to stretch his arms up, down or horizontally in place according to the given command when no target velocity is given, as shown in Figure~\ref{fig:defensive_locomotion}.
This leads to a goal state $\mathbf{g}_t^\text{loco} \in \mathbb{R}^4$ where three of the elements represent the target velocity in the form of direction and magnitude, like that for the dribbling policy, and an additional one for defensive pose indication.
The reward function is defined as:
\begin{equation}
    r_\text{loco} = \begin{cases}
        r_\text{nav} & \text{if $\mathbf{v}_\text{target}$ is given}, \\
        1 + 0.2\exp(-||\mathbf{v}||^2) + 0.8 r_\text{style} & \text{otherwise,}\\
    \end{cases}
\end{equation}
where $r_\text{nav}$ is the navigation reward from Eq.~\ref{eq:r_nav}.
When no target velocity is given, we want the character to stay in place by using the 2nd reward case to minimize its horizontal linear velocity $\mathbf{v}$.
Instead of using a conditional discriminator~\cite{dou2023c,tessler2023calm} for defensive stance control
we rely on the reward term $r_\text{style}$ to guide the stance styles for the policy.
$r_\text{style}$ is defined mainly according to to the arms' stretching state:
\begin{equation}
    r_\text{style} = \begin{cases}
       r_\text{block} & \text{if blocking command is given},\\
       r_\text{screen} & \text{if screening command is given},\\
       r_\text{defend} & \text{otherwise}.\\
    \end{cases}
\end{equation}
where
\begin{equation}\begin{split}
    r_\text{block} & = 0.25\max_\kappa \min\{1, \theta_\text{l}^\kappa/0.16\pi\} \\
     & \quad + 0.75 \max_\kappa \min\{1,(\theta_\text{u}^\kappa+0.212\pi)/0.376\pi\}, \\
    r_\text{screen} &  = 0.25 \min\{1, (0.5-||\mathbf{p}_\text{palm}^\text{left}-\mathbf{p}_\text{palm}^\text{right}||)/0.3\} \\
    & \quad + 0.75 \min_{i\in{l,u}} \min_\kappa \min\{1, (0.4\pi-\theta_\text{i}^\kappa)/0.8\pi\}, \\
    r_\text{defend} & = \min_\kappa \min\{1, (0.5\pi-|\theta_\text{u}^\kappa|)/0.334\pi)\}.
\end{split}
\end{equation}
Those reward measures the vertical angle for the upper ($\theta_\text{u}^\kappa$) and lower ($\theta_\text{l}^\kappa$) arms given $\kappa \in \{\text{left}, \text{right}\}$,
and encourages the character to reach the desired arm stretching pose.
For blocking,
the character obtains the maximal reward of $1$ if any of the upper arms and lower arms raise up more than $30^\circ$ at the same time.
For screening,
the character obtains the maximal reward when the two hands are close enough with a distance less than 0.2m, and all lower and upper arms put down with a vertical angle more than $70^\circ$.
In the third case,
the character obtains the maximal reward if all upper arms are stretched horizontally with a vertical angle within the range of $\pm 30^\circ$.

Because during training, the target velocity and arm stretching commands are generated randomly with a higher proportion of locomotion rather than staying in place,
to balance the learning of defending behaviors and locomotion, we start the training with all environments generating only zero target velocity for defensive behavior learning, and then progressively increase the proportion of environments accepting dynamic target velocity, which also could be zero.
This approach allows the character to master the simple defending poses first before learning locomotion.
Additionally, we adopt an adaptive sampling strategy to ensure the simulation steps spent on the learning of three different defending poses are equal roughly.
During interactive control,
the command of blocking and screening is given by the user,
while the default command for standing in place is the defensive stance.

\subsection{Foot Traveling Detection}\label{sec:traveling}

\begin{figure}
    \centering
    \includegraphics[width=.8\linewidth]{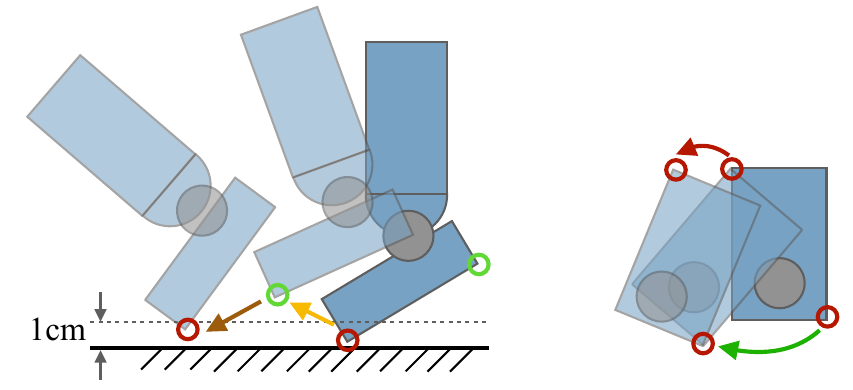}
    \caption{Demonstrations of foot traveling detection. Left: a foot with points contacting with ground (red circle) lifting up and dropping down would lead to traveling if the foot is a pivoting foot, or will let the other foot become the pivoting foot if the pivoting foot is undecided (two feet contacted with the ground at the same time previously when the ball is caught). Right: foot movement with at least one contacting point not changing is allowed (green arrow), but the traveling will be triggered if all contacting points move by a distance above 0.01m (red arrow) from the initial contacting points.}
    \label{fig:traveling_contact}
\end{figure}

\begin{algorithm}[t]
\SetKwInOut{Input}{Input}
\SetKwInOut{Output}{Output}
\Input{$p_{t-1} \in \{-1,0,1,2\}$: previous pivoting foot state, $p_0 =-1$.\newline
$\mathcal{P}, m_f, d_f, T_f^t, c_f^t$ from \textsc{FootMovementDetection} for $f \in \{\text{left foot}, \text{right foot}\}$.}

\Output{traveling $\in \{\text{true}, \text{false}\}$:
    traveling state.\newline
    $p_t$: updated pivoting foot state.
}

\nl traveling = false \\
\nl\uIf{ball is not held by any hand}{
    \nl $p_t=-1$ \tcp*{Undefined pivoting foot.}
    \nl $\mathcal{P}\leftarrow\emptyset$ \tcp*{Clear recorded contacting states.}
    \nl $c_f^t = \text{false}, \mathcal{T}_f^t = -4 \quad\forall f$
}
\nl\uElse{
    \nl $p_t = p_{t-1}$ \\
    \nl\uIf{$p_t = -1$}{
        \tcp{No foot-ground contact was detected.}
        \nl\uIf{$c_\text{left foot}^t$ and  $c_\text{right foot}^t$}{
        \tcp{Two feet contact ground simultaneously. Either foot can be the pivoting foot.}
        \nl    $p_t = 2$ 
        }
        \nl\uElseIf{$c_\text{left foot}^t$}{
        \nl    $p_t = 0$ \tcp*{Left foot pivots.}
        }
        \nl\uElseIf{$c_\text{right foot}^t$}{
        \nl    $p_t = 1$ \tcp*{Right foot pivots.}
        }
    }
    \nl\uElseIf{$p_t == 2$}{
        \tcp{Pivoting foot is not decided.}
        \tcp{Two feet contacted ground simultaneously after ball catching.}
    \nl    \uIf{$d_f$ for any $f \in \{\text{left foot}, \text{right foot}\}$}{
    \nl    traveling = true
    }
    \nl \uElseIf{$m_f$ for all $f \in \{\text{left foot}, \text{right foot}\}$}{
    \nl    traveling = true
    }
    \nl    \uElseIf{$c_\text{left foot}^t$ is false or ($c_\text{right foot}^t$ is true and $m_\text{left foot}$)}{
    \tcp{Left foot moves and right foot becomes the pivoting foot}
    \nl        $p_t = 1$\\
    \nl        traveling = $m_\text{right foot}$
        }
    \nl    \uElseIf{$c_\text{right foot}^t$ is false or ($c_\text{left foot}^t$ is true and $m_\text{right foot}$)}{
    \tcp{Right foot moves and left foot becomes the pivoting foot}
    \nl        $p_t = 0$\\
    \nl        traveling = $m_\text{left foot}$
        }
    }
    \nl\uElseIf{$p_t == 0$}{
        \tcp{Pivoting foot is the left foot.}
    \nl    traveling = $m_\text{left foot}$ or $d_\text{left foot}$
    }
    \nl\uElseIf{$p_t == 1$}{
        \tcp{Pivoting foot is the right foot.}
    \nl    traveling = $m_\text{right foot}$ or $d_\text{right foot}$
    }

    \uIf{$c_f$ and $T_f >t-4$ for all $f \in \{\text{left foot}, \text{right foot}\}$}{
    \tcp{4-frame tolerance to detect if two foot contact ground simultaneously.}
    \nl $p_t = 2$
    }
}
\caption{Pivoting Foot and Traveling Detection}
\label{alg:traveling}
\end{algorithm}

\begin{algorithm}[t]
\SetKwInOut{Input}{Input}
\SetKwInOut{Output}{Output}
\Input{$t$: the current timestep.\newline
$\mathcal{P}_{t-1}$: an array or table recording the contacting position of each foot link vertex $i$. $\mathcal{P}\leftarrow\emptyset$ during initialization.\newline
$T_f^{t-1}$ for $f \in \{\text{left foot}, \text{right foot}\}$: recording of the timestep at which the contact is detected for each foot link~$f$. $T_f = -4$ during initialization for a 4-frame tolerance to detect two-foot pivoting.\newline
$\mathbf{p}_i \in \mathbb{R}^3$: the current Cartesian coordinate of each foot link vertex $i$ in the global space. The third element, $\mathbf{p}_i[2]$, represents the vertical height above the ground.\newline
$c_f^{t-1} \in \{\text{true}, \text{false}\}$: the previous foot-ground contacting state.
}

\Output{$c_{f}^t \in \{\text{true}, \text{false}\}$: the ground contacting state of each foot.\newline
$m_{f} \in \{\text{true}, \text{false}\}$: the movement state of each foot. \newline
$d_{f} \in \{\text{true}, \text{false}\}$: the dropping state of each foot. \newline
$\mathcal{P}_t$: updated positions of contacting vertices.\newline
$T_f^t$: updated timestep recording of contacting foot.
}

\nl\For{each contacting vertex $i$ in $\mathcal{C}$}{
    \nl\uIf{$|\mathcal{P}[i][0]-\mathbf{p}_i[0]| > 0.01$ or $|\mathcal{P}[i][1]-\mathbf{p}_i[1]| > 0.01$}{
    \nl     vertex $i$ moves
    }
}

\nl\For{each $f \in \{\text{left foot}, \text{right foot}\}$}{
\nl $c_f = \text{false}$, $d_f = \text{false}$, $m_f = \text{false}$ \\
\nl\uIf{$\mathbf{p}_i[2] < 0.01$ for any vertex $i$ belongs to the foot $f$}{
    \nl $c_f^t = \text{true}$
    \tcp*{Foot $f$ is contacting ground.}
    \nl\uIf{$c_f^{t-1}$ is false}{
        \nl $d_f = \text{true}$ 
        \tcp*{Foot $f$ drops on ground.}
    }
    \nl\uIf{vertex $i$ moves for all $i$ belongs to the foot $f$}{
        \nl $m_f = \text{true}$
        \tcp*{Foot $f$ moves.}
    }
}
}

\nl $\mathcal{P}_t \leftarrow \mathcal{P}_{t-1}$, $T_f^t \leftarrow T_f^{t-1}$ \\
\nl\For{each $f \in \{\text{left foot}, \text{right foot}\}$}{
    \nl\For{each vertex $i$ belongs to the foot $f$}{
        \nl\uIf{$i$ not in $\mathcal{P}$ and $\mathbf{p}_i[2] < 0.01$}{
        \tcp{Record the first-time contacting coordinate for the vertex.}
        \nl     $\mathcal{P}_t[i]=\mathbf{p}_i$\\
        \nl\uIf{$T_f^t$ < 0}{
        \tcp{Record the first-time contacting timestep for the foot.}
        \nl    $T_f^t=t$
        }
        }
    }
}

\caption{Foot Movement Detection}
\label{alg:traveling2}
\end{algorithm}

We perform traveling detection according to the pivoting foot rules in real basketball games.
In practice, however, traveling is mainly decided visually.
It allows the player to rotate around a foot (the pivoting foot) rather than requiring that foot to be fixed on the ground strictly while holding the ball.
In our implementation, we introduce additional tolerance to avoid a too rigorous detection of traveling.
Given that each foot in our character model is shaped as a cuboid,
we consider the four vertices on the bottom side of the foot cube for contact detection.
A contact is recorded if any of the four vertices of a foot is less than 0.01m above the ground.
Traveling will be decided when the ball is holding and
if (1) a pivot foot falls back on the ground (from a non-contacting state to a contacting state) with the ball held by the hands,
or (2) a pivot foot moves with {\em all} recorded contact points traveling more than 0.01m horizontally.
The tolerance given in the second rule is crucial for allowing the character to perform pivoting where the body spins with one foot contacting the ground.
Figure~\ref{fig:traveling_contact} gives two examples of traveling detection.
In the right example, the movement along the green arrow is allowed, since one contacting point (the left top corner of the foot) is fixed and the movement is actually a rotation around that contacting point.
However, the movement along the red arrow, even after moving along the green arrow first, is considered traveling,
as we perform traveling detection by measuring the movement of the contacting points from the initial contact positions.

The pivoting foot rule in the basketball game treats the foot that first touches the ground as the pivoting foot when or after the ball is caught.
The rule allows the player to pivot on either foot if two feet are contacting the ground when catching the ball or touching the ground simultaneously after catching the ball.
In the latter case, once a foot lifts up, the other foot will become the pivoting foot.
However, it is too rigorous to perform the detection of two-foot pivoting at the time scale of 30Hz. 
Therefore, we introduce an additional 4-frame tolerance. 
If two feet contact the ground within 4 frames after the catch,
we consider that the two feet contact simultaneously, and either of the two feet can be treated as the pivoting foot (i.e. the undecided state of the pivoting foot).

For reference, we elaborate on our algorithm for pivoting foot and traveling detection in Algorithms~\ref{alg:traveling} and~\ref{alg:traveling2}.

\section{Transition Policy Training and Composing}

While the primitive policies can be trained independently at the same time,
our presented training scheme for policy transition (Type~B and~C in Figure~\ref{fig:main_method}) requires the preceding, succeeding, and intermediate (if there is) policies to be trained together to learn the transition.
In our implementation,
we run 512 simulation environments to train each primitive policy.
During the training for policy transition,
we run 512-by-$n$ environments at the same time where $n=2$ for mutual adaptation and $n=3$ for the scheme using intermediate policy.
Specially, for the rebounding policy, an additional batch of $512$ environments run in parallel to collect ball trajectories from which the the pre-rebounding ball state is extracted to initialize the ball for the character to rebound (cf. Section~\ref{sec:rebound}).
We perform policy composition for the case where the transition should be done automatically but the transitional states between two policies are ill-defined,
like the dribbling to shooting/passing transition through a gathering policy (Type C transitions), and the catching to shooting transition (Type B transitions).
During interactive control, for example,
we will call for the composed catching-to-shooting policy if catching and shooting commands are given at the same time.
If the command of catching and shooting is given sequentially, the system will perform catching using the catching-passing adapted policy first, and then, after receiving the shooting command, call the shooting-off-the-dribbling policy as the default executor of shooting. 

\begin{figure*}
    \centering
    \includegraphics[width=\linewidth]{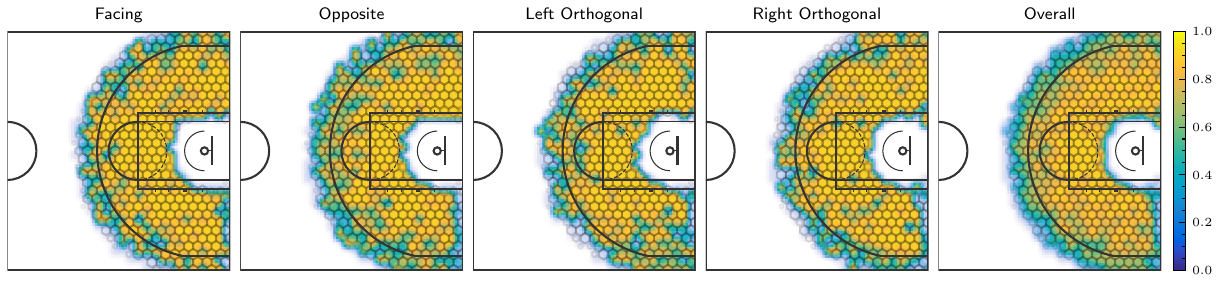}
    \caption{Heatmap of the shot percentage of our system %
    when the character dribbles at different approaching directions towards the hoop.}
    \label{fig:shot_percent_directions}
\end{figure*}

\begin{figure*}
    \centering
    \includegraphics[width=\linewidth]{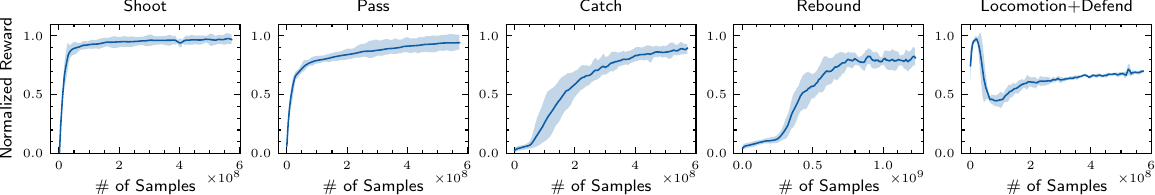}
    \caption{Curves of the primitive policy training performance. The shaded ranges show the performance over three training trials.}
    \label{fig:learning_curve_primitives}
\end{figure*}

\begin{figure}
    \centering
    \includegraphics[width=\linewidth]{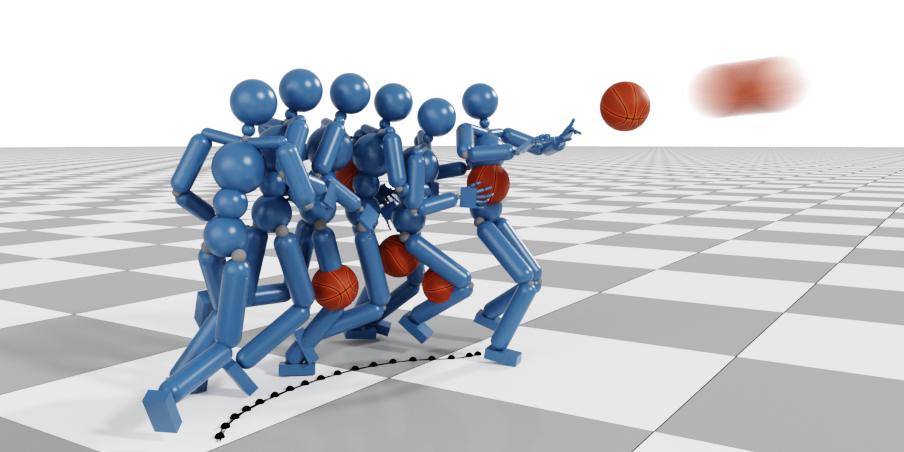}
    \caption{A demonstration of passing off the dribbling. Similar to the shooting-off-the-dribbling example, the introduced intermediate, gathering policy can adjust the body pose according to the passing target's position for seamless transition to passing.}
    \label{fig:add_off_demos}
\end{figure}

\section{Additional Results}
Figure~\ref{fig:shot_percent_directions} shows additional visualization results of the shot percentage of the policy trained by our system.
This figure is the visualization of the number reported in Table~\ref{tab:shot_percent}.
In Figure~\ref{fig:learning_curve_primitives}, we show the learning performance of the primitive policies, in terms of the task reward. The performance of the dribbling policy is shown in Figure~\ref{fig:dribble_rew}.
For the locomotion+defend case, the policy is trained for in-place defensive stances initially, and the ratio of locomotion training is increased gradually as the training goes on (cf. Section~\ref{sec:locomotion}). This leads to a performance drop at the early training stage with the introduction of locomotion, as shown in the right-most subplot in the figure.

\end{document}